\documentclass[fleqn]{article}
\usepackage{amsmath}
\usepackage{amsthm}
\usepackage{amssymb}
\usepackage{enumerate}
\usepackage{graphicx}
\newcommand{\qvec}[1]{\textbf{\textit{#1}}}
\newcommand{\beq}{\begin{equation}}
\newcommand{\eeq}{\end{equation}}
\newcommand{\bea}{\begin{eqnarray}}
\newcommand{\eea}{\end{eqnarray}}
\newcommand{\field}[1]{\mathbb{#1}}

\theoremstyle{plain}

\theoremstyle{definition}

\theoremstyle{remark}

\begin{document}

\vspace*{.1cm} \Large
\begin{center}
{\bf Grassmannian Connection Between Three- and Four-Qubit Observables, Mermin's Contextuality and Black Holes}
\end{center}
\large \vspace*{-.1cm}
\begin{center}
P\'eter L\'evay,$^{1}$ Michel Planat$^{2}$ and Metod Saniga$^{3}$

\end{center}
\vspace*{-.4cm} \normalsize
\begin{center}
$^{1}$Department of Theoretical Physics, Institute of Physics,
Budapest University of\\ Technology and Economics, H-1521
Budapest, Hungary

$^{2}$ Institut FEMTO-ST, CNRS, 32 Avenue de l'Observatorie,
F-25044 Besan\c con, France

$^{3}$Astronomical Institute, Slovak Academy of Sciences\\
SK-05960 Tatransk\' a Lomnica, Slovak Republic

\vspace*{.3cm} (10 August 2013)

\end{center}

\vspace*{-.3cm} \noindent \hrulefill

\vspace*{.1cm} \noindent {\bf Abstract}

\noindent
We invoke some ideas from finite geometry to map bijectively 135 heptads of mutually commuting {\it three}-qubit observables  into 135 symmetric {\it four}-qubit ones. After labeling the elements of the former set in terms of a seven-dimensional Clifford algebra, we present the bijective map and most pronounced actions of the associated symplectic group on both sets in explicit forms. 
This formalism is then employed to shed novel light on recently-discovered structural and cardinality properties of an aggregate of three-qubit Mermin's ``magic'' pentagrams. Moreover, some intriguing connections with the so-called black-hole--qubit correspondence are also pointed out.

\vspace*{.1cm} \noindent \hrulefill

\section{Introduction}
Generalized Pauli groups are widely used in the study of multipartite quantum systems associated with finite-dimensional Hilbert spaces. For $N$-partite systems these groups are built from $N$-fold tensor products of the familiar $2\times 2$ Pauli spin matrices and the $2\times 2$ identity matrix. Their principal applications lie within a rapidly evolving field of quantum information, where they are mainly related to quantum error correction codes \cite{Nielsen}. Such codes are constructed within the framework of the so-called stabilizer formalism \cite{Nielsen,Gottesman,Sloane}, making use of a simple fact that two elements (observables) in a Pauli group are either commuting, or anticommuting. This physically important property is  then encoded into the mathematical structure of a $2N$-dimensional vector space over the two-element field, endowed naturally with a symplectic structure.

Within the past few years, many important ramifications of this idea have appeared. In particular, it has been realized that the commutation algebra for $N$-qubit systems is encapsulated in a hierarchical structure of totally isotropic subspaces of this symplectic vector space, namely a symplectic polar space of rank $N$ and order two --- $\mathcal{W}(2N-1,\,2)$ \cite{Saniga1,HOS,Thas}. Because the sets of pairwise commuting operators are represented by such subspaces, this observation lends naturally itself to a finite geometric analysis of (various aggregates of) Mermin squares and Mermin's pentagrams \cite{SanigaLevay,SPPL,Holweck}, objects which furnish very economical proofs of Bell-Kochen-Specker-like theorems on hidden variables \cite{Mermin1,Mermin}.

Another interesting application of this idea concerns the recently-discovered Black-Hole--Qubit Correspondence (BHQC; for a recent comprehensive review, see \cite{BDL}). It has been observed that the structure of certain black hole entropy formulas, coming from charged extremal black hole solutions of effective supergravities stemming from toroidal compactifications of string theory, can elegantly be described by distinguished finite point-line incidence structures \cite{LSV,LSVP}. Truncations of these incidence structures to substructures called geometric hyperplanes (which in special cases form configurations like Mermin squares) have been found to correspond to truncations of the relevant supergravities featuring black hole solutions with a restricted set of charges. It turns out that the set of points of these incidence geometries can be mapped to the set of charges the particular supergravity model supports, and their sets of lines to the monomials in the entropy formula presented as a polynomial invariant \cite{LSVP}. Physically, the charges have their origin in wrapping configurations of extended objects (membranes) on special submanifolds of the extra-dimensions \cite{Becker}. It has also been demonstrated \cite{LSVP} that for a physically consistent realization of charges in a finite-geometric context, these incidence structures should be labeled in a noncommutative manner by elements of the generalized Pauli group for three-qubits. Moreover, since different string theories are connected by a web of dualities \cite{Becker}, it may happen that a particular labeling within a framework suggested by one particular string theory should be replaced by another labeling, suggested by its dual counterpart. Hence, such dualities in string theories strongly hint at alternative realizations of the same finite-geometric structures. Thus,  on the one hand, one can  map the point-set of an incidence geometry to the set of observables and its line-set to tuples of pairwise commuting observables. On the other hand, one can equally well adopt a dual view, regarding a {\it single point} as a {\it set} of mutually commuting observables, and a line as a tuple of such sets whose pairwise overlapping obeys a prescribed set of rules. A pattern very similar to this has already appeared \cite{Levfano} in an investigation of the structure of the $E_{7(7)}$-symmetric black hole entropy formula of $N=8$ supergravity, based on the incidence structure of the Fano plane.

Surprisingly, this idea has also emerged in investigations of the Bell and Bell-Kochen-Specker theorems on hidden variables. There, it is usually emphasized that different sets of mutually
commuting observables can be interpreted as representatives of physical situations associated with different possible experimental arrangements/set-ups. If in a hypothetical hidden-variables-theory the result of an observation depends not only on the state of the system, but also on the complete disposition of the apparatus, the corresponding theory is called {\it contextual}.
It has already been established (see, for example, \cite{SanigaLevay,SPPL,Holweck} and references therein) that our finite geometries also provide a promising formal footing for dealing with contextuality and associated ``context spaces.'' 

A major theme of this paper is the study of the structure of such context space for three-qubits. Our reason for deciding to conduct a detailed investigation of this special case is as follows. First, for three-qubits the invariance group of the symplectic form which governs the commutation structures of the corresponding operators is $Sp(6,2)$. As observed by one of us \cite{Sole}, and elaborated in more detail by \cite{Geemen}, this group is related to the Weyl group of the exceptional group $E_7$, $W(E_7)$, as $W(E_7)/\field{Z}_2=Sp(6,2)$. $W(E_7)$ is the physically important subgroup representing electric-magnetic duality inside the full duality group for toroidal compactifications of string- and $M$-theories to four dimensions \cite{Pioline}. Hence, a clear
understanding of representations of this group on objects like our context space can foster a deeper understanding of the BHQC. Second, the elements of context space for three-qubits are heptads of pairwise commuting operators, whose total number amounts to 135. From these heptads one can extract three- and four-tuples of observables that represent, respectively, basic building blocks of Mermin squares and Mermin pentagrams. In order to have a deeper, computer-free understanding of recent observations made on such objects \cite{SanigaLevay,SPPL,Holweck,LSVP}, as a first step it is vital to present a broader finite-geometric setting for these structures.

Since the main aim of the present paper is to set such a finite geometric ground for further applications in connection with the BHQC and more general issues of quantum contextuality, we shall also give an explicit form of the bijection between the three-qubit context space, viz. the above-mentioned set of 135 heptads of pairwise commuting observables, and the set of 135 symmetric four-qubit operators. In a finite geometric language, this is a bijection between the 135 maximal totally isotropic subspaces of the polar space $\mathcal{W}(5,2)$ and the 135 points of the hyperbolic quadric $\mathcal{Q}^{+}(7,2)$ fully embedded in the polar space $\mathcal{W}(7,2)$. Though this mapping is well known in the mathematics literature \cite{Cossking1,Gow,Cossidente}, mainly as the spin embedding or the spin module for the group $Sp(6,2)$, its explicit form --- to the best of our knowledge --- has neither been worked out in detail, nor employed in (mathematical) physics as yet. From the physical point of view,  this bijection may in the future play a role very similar to that of the famous Klein correspondence (see, e.\,g., \cite{Hirschfeld}), which over the complex numbers has already been used in twistor quantization methods of space-time geometry \cite{Penrose,Ward}.

The paper is organized as follows. Section 2 sets the stage for our investigations by summarizing the basics of the finite-geometric background needed later on. Although this
summary will be presented  merely for the three-qubit case, it generalizes trivially to $N$-qubits. In Section 3, in the spirit of \cite{Shaw1,Shaw2}, we label the 135 heptads (whose set will be denoted by $\mathcal{I}$ and referred to as the context space in the sequel) by the elements of a seven-dimensional Clifford algebra. Here, the action of the associated symplectic group $Sp(6,2)$ on this context space is also discussed. In Section 4, the context space is related to the space of separable trivectors satisfying  Pl\"ucker relations and an additional constraint demanding compatibility with the symplectic polarity. The trivectors arising in this way and encapsulating information on $\mathcal{I}$ are the {\it primitive trivectors}. In Section 5, we endow the binary $14$-dimensional vector space of primitive trivectors with a symplectic form and show that this space can naturally be expressed as a direct sum of an $8$- and a $6$-dimensional
vector space over $\field{Z}_2$. We will further demonstrate that  the $8$-dimensional space can be identified with the space of real four-qubit observables. Since the latter space comprises exactly $135$ symmetric guys,  a bijection will be established between these and the $135$ heptads of the context space. Employing this bijection, we will subsequently calculate the
irreducible \cite{Cossking1,Cossidente,Gow} action of $Sp(6,2)$ on the space of four-qubit observables. In Section 6, we shall first apply our formalism to get deeper insights into the nature of the space of Mermin's pentagrams and to furnish an elegant, computer-fee explanation of some recent findings in this respect \cite{SanigaLevay,Holweck}. In Section 7, as a second interesting application of our formalism,  by reiterating an observation of \cite{Geemen} we shall show that our space of contexts can be related to the set of possible embeddings of seven copies of the three-qubit SLOCC group \cite{Dur} inside the exceptional group $E_7$ --- an idea that has originally been introduced within the context of the BHQC \cite{Ferrara,Levfano}. Finally, Section 8 is reserved for concluding remarks.

\section{The geometry of the three-qubit Pauli group}

Let us consider the set of three-qubit observables acting on the
state space for three-qubits, ${\mathcal H}\equiv
\field{C}^2\otimes \field{C}^2\otimes \field{C}^2$. Such
observables are of the form $A_1\otimes A_2\otimes A_3$, where the
operators $A_1,A_2,A_3$ are elements from the set $\{\pm
I,\pm\sigma_x,\pm\sigma_y,\pm\sigma_y\}\equiv\{\pm I,\pm X,\mp
iY,\pm Z\}$. Here, $\sigma_x,\sigma_y,\sigma_z$ are the standard
$2\times 2$ Pauli spin matrices and $I$ is the $2\times 2$
identity matrix. In what follows, we shall consider instead the real
operators of the form $\mathcal{A}_1\otimes\mathcal{A}_2\otimes
\mathcal{A}_3$, where $\mathcal{A}_1,\mathcal{A}_2,\mathcal{A}_3$
are elements from the set  $ \mathcal{P}_1\equiv\{\pm I,\pm X,\pm
Y,\pm Z\}$. $\mathcal{P}_1$ is a group and will be called the
Pauli group for a single qubit. Notice that the operators $I,X,Z$
are real and symmetric, and the operator $Y$ is real and antisymmetric.
Elements of the form $\mathcal{A}_1\otimes\mathcal{A}_2\otimes
\mathcal{A}_3$, where $\mathcal{A}_1,\mathcal{A}_2,
\mathcal{A}_3\in\mathcal{ P}_1$, are elements of $\mathcal{P}_3$,
the three-qubit Pauli group. (The following considerations can be
straightforwardly generalized to $N$ qubits.)

An arbitrary element $x$ of $\mathcal{P}_3$ can be written in the
form 
\beq x=(-1)^s(Z^{a_1}X^{b_1}\otimes Z^{a_2}X^{b_2}\otimes
Z^{a_3}X^{b_3})\equiv(s,v)=(s,a_1,b_1,a_2,b_2,a_3,b_3)\in\mathcal{P}_3,
\label{kanform} 
\eeq 
where each superscript can acquire two values, 0 and 1.
\noindent The product of two elements
$x,x^{\prime}\in \mathcal{P}_3$ is 
\beq
xx^{\prime}=(s+s^{\prime}+\sum_{i=1}^3a_i^{\prime}b_i,a_1+a_1^{\prime},\dots,
b_3+b_3^{\prime}). \label{prod} \eeq 
\noindent Hence, two elements of $\mathcal{P}_3$
commute if, and only if, 
\beq
\sum_{i=1}^{3}(a_ib_i^{\prime}+a_i^{\prime}b_i)=0. \label{komm}
\eeq 
The commutator subgroup of $\mathcal{P}_3$ coincides with its
center $Z(\mathcal{P}) = \{I\otimes I\otimes I,-I\otimes
I\otimes I\}$; hence, the central quotient
$V_3=\mathcal{P}_3/Z(\mathcal{P}_3)$ is an Abelian group which --- by
virtue of (\ref{prod}) --- is also a six-dimensional vector space over
$\field{Z}_2$, i.\,e. $V_3\equiv \field{Z}_2^6$. Moreover, 
the left-hand-side of (\ref{komm}) defines on $V_3$ a symplectic form\beq
\langle\cdot,\cdot\rangle:V_3\times V_3\to \field{Z}_2,\qquad
(v,v^{\prime})\mapsto \langle v,v^{\prime}\rangle\equiv
\sum_{i=1}^3 (a_ib_i^{\prime}+b_ia_i^{\prime}).\label{ezittaszimpl} \eeq \noindent The
elements of the vector space $(V_3,\langle\cdot,\cdot\rangle)$ are
equivalence classes corresponding to pairs of the form
$\{\mathcal{A}_1\otimes\mathcal{A}_2\otimes \mathcal{A}_3,
-\mathcal{A}_1\otimes\mathcal{A}_2\otimes \mathcal{A}_3\}$, i.\,e.
they are three-qubit operators defined {\it up to a sign}. In the
sequel, we shall employ a short-hand notation
$\mathcal{A}_1\mathcal{A}_2\mathcal{A}_3 \equiv \{\mathcal{A}_1\otimes\mathcal{A}_2\otimes \mathcal{A}_3,
-\mathcal{A}_1\otimes\mathcal{A}_2\otimes \mathcal{A}_3\}$. Alternatively, we will also refer to
this object as an element $v$ of $(V_3,\langle\cdot,\cdot\rangle)$.

Since any single-qubit operator $\mathcal{A}$ can be written, sign disregarded, in the
form $\mathcal{ A}=Z^aX^b$, where $a,b\in \field{Z}_2$, one can associate with it a two component vector $(a,b)\in
V_1\equiv\field{Z}_2^2$. Hence, we have \beq I\mapsto (00),\qquad
X\mapsto (01),\qquad Y\mapsto (11),\qquad Z\mapsto (10).
\label{corr1} \eeq For a three-qubit operator 
we adopt the following ordering convention \beq
\mathcal{A}_1\mathcal{A}_2\mathcal{A}_3\leftrightarrow
(a_1,a_2,a_3,b_1,b_2,b_3)\in V_3. \label{konvencio} \eeq \noindent
Hence, for example, 
\beq \{X\otimes Y\otimes Z,-X\otimes Y\otimes
Z\}\leftrightarrow XYZ\leftrightarrow (0,1,1,1,1,0) \label{pelda}
\eeq 
\noindent and the canonical basis vectors in $V_3$ are
associated to three-qubit operators as follows \beq
ZII\leftrightarrow e_1=(1,0,0,0,0,0), \ldots,
IIX\leftrightarrow e_6=(0,0,0,0,0,1). \label{bazis} \eeq \noindent
With respect to this basis, the matrix of the symplectic form
is
\beq J_{\mu\nu}\equiv \langle e_{\mu},e_{\nu}\rangle
=\begin{pmatrix}0&0&0&1&0&0\\0&0&0&0&1&0
\\0&0&0&0&0&1\\
        1&0&0&0&0&0
    \\  0&1&0&0&0&0
    \\  0&0&1&0&0&0\end{pmatrix}, \qquad \mu,\nu=1,2,\ldots, 6.
        \label{simplmatr}
        \eeq
        \noindent
Since the dimensionality of $V_3$ is even and the symplectic form is
non-degenerate, the group preserving this symplectic polarity is isomorphic to the
group $Sp(6,2)$. This group acts on the row
vectors of $V_3$ via $6\times 6$ matrices $S\in Sp(6,2)$ from the
right, leaving the matrix $J$ invariant \beq
v\mapsto vS,\qquad SJS^t=J. \label{transz} \eeq \noindent It is
known that $\vert Sp(6,2)\vert =1451520=2^9\cdot 3^4\cdot 5\cdot
7$ and that this group is generated by transvections \cite{Geemen}
$T_w\in Sp(6,2), w\in V_3$ of the form \beq T_w:V_3\to V_3,\qquad
v\mapsto T_wv=v+\langle v,w\rangle w, \label{transvections} \eeq
\noindent which is indeed symplectic, 
\beq \langle
T_wu,T_wv\rangle=\langle u,v\rangle. \label{szimpltulajd} 
\eeq

\noindent 
There is a surjective homomorphism from $W(E_7)$ to $Sp(6,2)$ with kernel $\field{Z}_2$. 
This homomorphism provides an interesting link between exceptional groups and three-qubit systems. Within the context of Quantum Information this intriguing link was first noticed by one of us \cite{Sole}, and it is also related to the Black-Hole--Qubit Correspondence \cite{Ferrara,Levfano}. A particularly nice elaboration of this homomorphism can be found in the paper of Cerchiai and van Geemen \cite{Geemen}.

Given an element $v\in V_3$, let us define the quadratic form \beq
Q_0\equiv \sum_{i=1}^3 a_ib_i. \label{kankvadrat} 
\eeq \noindent
It is easy to check that for vectors representing symmetric
operators $Q_0(v)=0$ and for antisymmetric ones $Q_0(v)=1$;
moreover, 
\beq \langle u,v\rangle =Q_0(u+v)+Q_0(u)+Q_0(v).
\label{kvdratosszef} 
\eeq 
\noindent This quadratic form can be
regarded as the one labeled by the trivial element of $V_3$,
corresponding to the trivial observable $III$. There are, however,
$63$ other quadratic forms $Q_w$ associated with the symplectic
form $\langle\cdot\vert\cdot\rangle$ labeled by $63$ nontrivial
elements of $V_3$ \beq \langle u,v\rangle =Q_w(u+v)+Q_w(u)+Q_w(v);
\label{kvdratosszef2} \eeq \noindent they are defined as \beq
Q_w(v)=Q_0(v)+\langle w,v\rangle^2,
\label{ujkvadforms}\eeq\noindent where the square can be omitted as we work over $\field{Z}_2$. For more details on these
quadratic forms, we refer the reader to \cite{Shaw2,Levvran}; here,
we merely note that a form labeled by a symmetric
observable ($Q_0(w)=0$) represents in the associated projective space $PG(5,2)$ a locus of points satisfying $Q_w(v)=0$,
which is also called a {\it hyperbolic} quadric and usually denoted by $\mathcal{Q}^+(5,2)$. 
On the other hand, the locus of points satisfying $Q_w(v)=0$ where $Q_0(w)=1$
is denoted by $Q^-(5,2)$ and called an {\it elliptic} quadric.
These quadrics can trivially be generalized to $N$-qubits. Thus, for
example, a hyperbolic quadric in $PG(7,2)$, $\mathcal{Q}^+(7,2)$,
given by the locus $Q_0(v)=0$, $v\in V_4$, is just the quadric whose
points correspond to the symmetric four-qubit observables.

By virtue of the special character of the field $\field{Z}_2$, a one-dimensional subspace of $V_3$ (consisting of elements of the form $\lambda v$ where $\lambda\in\field{Z}_2$ and $v\in V_3$) is spanned by a unique nonzero vector $v\in V_3$. Hence, the 63 points of the projective space $PG(5,2)$ can be identified with the $63$ nonzero vectors of $V_3$. Since the vector space $V_3$
underlying $PG(5,2)$ is equipped with a symplectic form, one can determine on any subspace $W$ of $PG(5,2)$ a symplectic polarity $\perp$, i.\,e. the map 
\beq W\mapsto W^{\perp}, \quad {\rm where}\quad
W^{\perp}=\{v\in V_3\vert wJv^T=0, \forall w\in
W\},\label{polarity}\eeq \noindent
and call such subspace {\it non-isotropic},
{\it isotropic} or {\it totally isotropic} according as $W\cap W^{\perp}=\{0\}$, $W\cap W^{\perp}\neq \{0\}$
or $W\subset W^{\perp}$, respectively. A maximal totally isotropic subspace is isomorphic to $PG(2,2)$ --- the Fano plane. The space of all totally isotropic subspaces of  $(PG(5,2),\perp)$
is the symplectic polar space of rank three and order two,  $\mathcal{W}(5,2)$; it contains 63 points, 315 lines and 135 planes (see, e.\,g., \cite{pralle}).

Let us illustrate these abstract concepts in terms of the
physically relevant  structures of three-qubit operators (defined
up to a sign). The $63$ points of
$PG(5,2)$ (as well as of $\mathcal{W}(5,2)$) are all 63 nontrivial operators of $\mathcal{P}_3$.
$W$ is any subset of pairwise commuting operators and
$W^{\perp}$ is the set of operators {\it commuting} with each member of this particular subset. A line $L\in PG(5,2)$ is an object
of the form $L=\lambda v+\mu u$, where $u,v\in PG(5,2)$ and
$\lambda,\mu\in \field{Z}_2$. It contains the following three points
$u,v,u+v$ and thus corresponds to a triple of operators
such that the product of any two of them yields the
third one. A line (as well as a plane) of $PG(5,2)$ belongs to $\mathcal{W}(5,2)$ if, and only if, the
corresponding operators are also {\it pairwise commuting}. 
A plane of $\mathcal{W}(5,2)$ represents a heptad, i.\,e. a maximum set, of mutually commuting three-qubit observables.
An illustrative example is furnished by the heptad
$\{IIX,IXI,IXX,XII,XIX,XXI,XXX\}$;  the seven lines of this plane have the following representatives
\beq \{IIX,IXI,IXX\},~\{IIX,XII,XIX\},~\{IIX,XXI,XXX\},~\{IXI,XII,XXI\},
\nonumber \eeq
\noindent \beq 
\{XXX,XII,IXX\},~\{XXI,XIX,IXX\},~\{XXX,XIX,IXI\}.\label{linesetfano1}\eeq\noindent
The central object of our reasoning in the subsequent sections will be the set of 135 heptads of mutually commuting elements of $\mathcal{P}_3$, {\it aka} the set of 135 planes of $\mathcal{W}(5,2)$.\footnote{By a slight abuse of notation, we shall use the same symbol, $\mathcal{I}$, for both the sets.}  
Before proceeding further, and given the fact that the language of finite geometry employed in the sequel is not a commonplace in either quantum information or high energy physics,  
we think the reader finds convenient a short summary of basic cardinality characteristics of key finite geometrical objects involved:\\ \\ 

\begin{tabular}{lrr}
\hline \hline
\vspace*{-.3cm} \\
Name of Object                    & No. of Points    & No. of Maximal Subspaces   \\
\hline
\vspace*{-.3cm} \\
Projective (Fano) plane PG(2,\,2) & 7 & 7 \\
Projective space PG(3,\,2) & 15 & 15 \\
Hyperbolic (Klein) quadric   $\mathcal{Q}^+(5,2)$   & 35         & $2 \times 15 = 30$ \\
Symplectic polar space $\mathcal{W}(5,2)$   & 63         & 135 \\
Hyperbolic quadric   $\mathcal{Q}^+(7,2)$   & 135        & $2 \times 135 = 270$ \\
Symplectic polar space $\mathcal{W}(7,2)$   & 255        & 2295 \\
Split Cayley hexagon of order two           & 63         & 63  \\

\vspace*{-.3cm} \\
\hline \hline
\end{tabular}
 
\vspace*{.5cm}

\section{Clifford labeling of the context space}

Our next goal is to find a convenient labeling of the elements of $\mathcal{I}$ and then describe the action of $Sp(6,2)$ on this set. To begin with, we first label the 63 nontrivial
three-qubit operators in terms of the generators of a Cliff(7) algebra 
\beq \{\Gamma_a,\Gamma_b\}=-2\delta_{ab},\qquad
a,b=1,2,3,4,5,6,7. \label{cliff7}\eeq \noindent A possible choice
of generators satisfying Eq. (\ref{cliff7}) is \beq
\Gamma_1=I\otimes I\otimes Y,\quad \Gamma_2=Z\otimes Y\otimes
X,\quad \Gamma_3=Y\otimes I\otimes X,\quad \Gamma_4=Y\otimes
Z\otimes Z, \nonumber\eeq\noindent \beq \Gamma_5=X\otimes Y\otimes
X,\quad \Gamma_6=I\otimes Y\otimes Z,\quad \Gamma_7=Y\otimes
X\otimes Z. \label{explicitclifford}\eeq \noindent Assuming that $1\leq
a<b<c< \cdots <e<f\leq 7$, we will use the shorthand notation
$\pm\Gamma_a\Gamma_b\Gamma_c \cdots \Gamma_e\Gamma_f\equiv
abc \cdots ef$. When needed, for products we will sometimes use a
cyclic reordering (e.\,g., ``$267$'' $\rightarrow$ ``$672$''). Notice
that antisymmetric operators can be represented by singlets and
doublets of Clifford generators, whereas symmetric ones
are expressed in terms of triplets; thus, for example,
$\Gamma_1\Gamma_2=Z\otimes Y\otimes Z$ is an antisymmetric
operator and $\Gamma_1\Gamma_2\Gamma_5=Y\otimes
I\otimes Y$ is a symmetric one.

In a previous paper \cite{LSV} we have shown that there is an
automorphism of order seven which acts on $\mathcal{P}_3$ via
conjugation. Under this automorphism the $63$ nontrivial
three-qubit operators split into nine different orbits containing seven
operators each. One of the orbits consists of the
(\ref{explicitclifford}) basis vectors of the Clifford algebra. In
order to see this, let $\alpha\equiv(1,2,\cdots, 7)$ denote
the permutation $1\mapsto 2\mapsto \cdots \mapsto 7\mapsto 1$. Then
the matrix  $\mathcal{D}(\alpha)$ that acts via
conjugation \beq
\mathcal{A}_1\otimes\mathcal{A}_2\otimes\mathcal{A}_3\mapsto
\mathcal{D}^{-1}(\alpha)
(\mathcal{A}_1\otimes\mathcal{A}_2\otimes\mathcal{A}_3)\mathcal{D}(\alpha)\label{conjugateaction}
\eeq\noindent and shifts cyclically the generators of our
Clifford algebra, i.\,e.
$\Gamma_1 \mapsto \Gamma_2 \mapsto \cdots \mapsto \Gamma_7 \mapsto \Gamma_1$,
is of the form \cite{LSV}\beq {\cal
D}(\alpha)\equiv\begin{pmatrix}P&Q&0&0\\0&0&Q&P\\0&0&QX&PX\\PX&QX&0&0\end{pmatrix},\quad
P=\begin{pmatrix}1&0\\0&0\end{pmatrix},\quad
Q=\begin{pmatrix}0&0\\0&1\end{pmatrix}. \nonumber \eeq \noindent
An alternative form of this $8\times 8$ matrix can be
given \cite{LSV} in terms of the two-qubit CNOT operations familiar
from Quantum Information,
 \beq \mathcal
{D}(\alpha)=(C_{12}C_{21})(C_{12}C_{31})C_{23}(C_{12}C_{31}),
\nonumber \eeq where\beq
C_{12}=\begin{pmatrix}I&0&0&0\\0&I&0&0\\0&0&0&I\\0&0&I&0\end{pmatrix},\qquad
C_{21}=\begin{pmatrix}I&0&0&0\\0&0&0&I\\0&0&I&0\\0&I&0&0\end{pmatrix},
\nonumber \eeq \beq
C_{23}=\begin{pmatrix}I&0&0&0\\0&X&0&0\\0&0&I&0\\0&0&0&X\end{pmatrix},
\qquad
C_{31}=\begin{pmatrix}P&0&Q&0\\0&P&0&Q\\Q&0&P&0\\0&Q&0&P\end{pmatrix}.
\label{cnots2} \nonumber \eeq
\noindent
As every three-qubit operator can be expressed in terms of the basis vectors
(\ref{explicitclifford})  of the Clifford algebra, the nine orbits under $\mathcal{D}(\alpha)$ explicitly read
 \beq (1,2,3,4,5,6,7)\leftrightarrow
(IIY,ZYX,YIX,YZZ,XYX,IYZ,YXZ), \label{clifforbit}\eeq \noindent
\beq (12,23,34,45,56,67,71)\leftrightarrow
(ZYZ,XYI,IZY,ZXY,XIY,YZI,YXX), \eeq \noindent \beq
(13,24,35,46,57,61,72)\leftrightarrow
(YIZ,XXY,ZYI,YXI,ZZY,IYX,XZY), \eeq \noindent \beq
(14,25,36,47,51,62,73)\leftrightarrow
(YZX,YII,YYY,IYI,XYZ,ZIY,IXY), \eeq \noindent \beq
(123,234,345,456,567,671,712),\leftrightarrow
(XYY,ZXZ,XXZ,ZZX,ZXX,YZY,XZI), \eeq \noindent \beq
(125,236,347,451,562,673,714),\leftrightarrow
(YIY,XIZ,YYX,ZXI,YYZ,IZX,IYY), \eeq \noindent \beq
(135,246,357,461,572,613,724),\leftrightarrow
(ZYY,XZX,XZZ,YXY,IXZ,YYI,ZIX), \label{lastorbit1}\eeq \noindent \beq
(124,235,346,457,561,672,713),\leftrightarrow
(XXI,IIX,IXX,XIX,XII,XXX,IXI), \label{xesek}\eeq \noindent \beq
(126,237,341,452,563,674,715),\leftrightarrow
(ZII,ZZZ,IZI,IZZ,ZIZ,IIZ,ZZI). \label{zsek}\eeq 
\noindent Moreover, $\mathcal{D}(\alpha)$ belongs to $SO(8)$ and is orthogonal, $\mathcal{D}^T(\alpha)=\mathcal{D}^{-1}(\alpha)$; hence, under
its conjugate action symmetric operators are mapped to symmetric,
and antisymmetric operators to antisymmetric ones. We have
four orbits of antisymmetric operators and five orbits of symmetric
ones.

Let us also give the $6 \times 6$ matrix representation $D(\alpha)$
of the cyclic shift $\alpha=(1234567)$. According to
Eq. (\ref{transz}), $D(\alpha)$ acts from the right on the
elements of $v\in V_3=\field{Z}_2^{6}$ regarded as six-component
{\it row} vectors with \beq
D(\alpha)=\begin{pmatrix}K&0\\0&L\end{pmatrix}, \quad
K=\begin{pmatrix}1&1&1&\\0&1&1\\1&1&0\end{pmatrix},\quad
L=\begin{pmatrix}1&1&1\\1&1&0\\0&1&1\end{pmatrix},\label{dalfa}
\eeq \noindent where off-diagonal blocks contain merely
zeros. Clearly, $D(\alpha)JD^T(\alpha)=J$ and $D(\alpha)^7=1$, which implies that
$D(\alpha)\in Sp(6,2)$ and is of order seven. For example, the
action of $D(\alpha)$ yields 
\beq (001001)\mapsto (110011)\mapsto
(100101)\mapsto(111100)\mapsto (010111)\mapsto(011010)\mapsto
(101110), \eeq 
\noindent which corresponds to the orbit given by
Eq.\,(\ref{clifforbit}).

We already know that $Sp(6,2)$ is generated by symplectic
transvections. However, for the reader's convenience, we shall also give its
presentation featuring merely two generators $\alpha$ and
$\beta$, one of them being our cyclic shift of order seven and
the other a particular element of order two. The presentation in question is \cite{Vrana} 
\beq
Sp(6,2)=\langle\alpha,\beta\vert\alpha^7=\beta^2=(\beta\alpha)^9=
(\beta\alpha^2)^{12}=[\beta,\alpha]^3=[\beta,\alpha^2]^2=1\rangle.
\label{presentsympl} \eeq 
\noindent 
The $6\times 6$ matrix
representation $D(\beta)$ of the generator $\beta$, which acts on $V_3$
from the right, is given by \beq
D(\beta)=\begin{pmatrix}1&0&0&0&0&0\\0&1&0&0&0&0\\1&1&1&0&0&1\\1&1&0&1&0&1\\
1&1&0&0&1&1\\0&0&0&0&0&1\end{pmatrix}.\label{betamatrix}\eeq This
matrix is again symplectic, $D(\beta)JD^T(\beta)=J$, and of order
two. The action of $D(\beta)$ induced on
three-qubit operators, defined up to a sign,  has the following form \beq
IIZ\leftrightarrow ZZY,\quad IIY\leftrightarrow ZZZ,\quad
IXI\leftrightarrow ZYX,\quad IXX\leftrightarrow ZYI, \eeq \noindent
 \beq
IZZ\leftrightarrow ZIY,\quad IZY\leftrightarrow ZIZ,\quad
IYI\leftrightarrow ZXX,\quad IYX\leftrightarrow ZXI, \eeq \noindent
\beq XII\leftrightarrow YZX,\quad XIX\leftrightarrow YZI,\quad
XXZ\leftrightarrow YYY,\quad XXY\leftrightarrow YYZ, \eeq \noindent
\beq XZI\leftrightarrow YIX,\quad XZX\leftrightarrow YII,\quad
XYZ\leftrightarrow YXY,\quad YXZ\leftrightarrow XYY, \eeq \noindent
with the remaining elements being left invariant. One observes that
the transformations above can be obtained via (up to a sign) multiplication by the operator $ZZX$, which anticommutes with all of the operators
appearing in the list; the remaining $31$ invariant operators are precisely those that commute with $ZZX$.
Hence, according to Eq.\,(\ref{transvections}), $D(\beta)$ is just a
matrix representative of the transvection defined by
$v=(110001)\leftrightarrow ZZX$. Notice also that, unlike $\alpha$,
the generator $\beta$ cannot be lifted to a conjugate action on
$\mathcal{P}_3$ of the (\ref{conjugateaction}) type via an {\it
orthogonal} matrix. This is immediately obvious from the fact that
$D(\beta)$ maps symmetric operators to antisymmetric ones,
and vice versa. We also mention that in terms of the labels
referring to our $Cliff(7)$ algebra, the action of $\beta$ can be
summarized as 
\beq (123,7),(237,1),(137,2),(127,3),\eeq
\noindent
\beq (14,156),(15,146),(16,145), \qquad (24,256),(25,246),(26,245),\eeq\noindent \beq
(34,356),(35,346),(36,345), \qquad (47,567),(57,467),(67,457).\eeq\noindent
The above-described $Cliff(7)$-labeling of three-qubit operators, although being of importance of its own, also leads to a neat description of the context space, $\mathcal{I}$, with two classes of elements of cardinality 105 and 30. 

A Fano plane of the first class is of the type $\{7,12,34,56,127,347,567\}$, i.\,e. it comprises four antisymmetric and three  symmetric operators. The pattern clearly shows that the corresponding seven operators are pairwise commuting. In order to also understand
the structure of its lines, one notes that $1234567\leftrightarrow III$,
and $\Gamma_a^2\leftrightarrow III$. Hence, the double occurrence
of any number, as well as the occurrence of all numbers from $1$ to
$7$,  yields the identity. The lines of the Fano plane
are thus the combinations $(7,12,127), (7,34,347), (7,56,567)$, and the
ones $(127,347,567), (12,34,567),(12,56,347),(34,56,127)$. Notice
that the three antisymmetric operators $12,34,56$ already determine
the Fano plane; indeed, they cannot be collinear since
$123456\leftrightarrow 7$.  As a consequence, all the planes featuring the operator $7\leftrightarrow YXZ$ can be characterized
by all the disjoint triples of doublets featuring all the numbers
from $1$ to $6$. There are $15$ doublets, of which $15$
such triples can be formed. These $15$ triples form the lines of a
$PG(3,2)$. Hence, there are $15$ planes featuring the operator
$YXZ$ related to this specific $PG(3,2)$:
\beq \{7,16,25,34,167,257,347\}
\leftrightarrow\{YXZ,IYX,YII,IZY,YZY,IXZ,YYX\},\label{C1}\eeq\noindent
\beq\{7,16,24,35,167,247,357\}
\leftrightarrow\{YXZ,IYX,XXY,ZYI,YZY,ZIX,XZZ\},\label{C2}\eeq\noindent
\beq\{7,12,34,56,127,347,567\}
\leftrightarrow\{YXZ,ZYZ,IZY,XIY,XZI,YYX,ZXX\},\label{C3}\eeq\noindent
\beq\{7,14,25,36,147,257,367\}
\leftrightarrow\{YXZ,YZX,YII,YYY,IYY,IXZ,IZX\},\label{C4}\eeq\noindent
\beq\{7,16,23,45,167,237,457\}
\leftrightarrow\{YXZ,IYX,XYI,ZXY,YZY,ZZZ,XIX\},\label{C5}\eeq\noindent
\beq \{7,13,25,46,137,257,467\}
\leftrightarrow\{YXZ,YIZ,YII,YXI,IXI,IXZ,IIZ\}, \eeq\noindent \beq
\{7,15,26,34,157,267,347\}
\leftrightarrow\{YXZ,XYZ,ZIY,IZY,ZZI,XXX,YYX\}, \eeq\noindent \beq
\{7,14,26,35,147,267,357\}
\leftrightarrow\{YXZ,YZX,ZIY,ZYI,IYY,XXX,XZZ\}, \eeq\noindent \beq
\{7,12,45,36,127,457,367\}
\leftrightarrow\{YXZ,ZYZ,ZXY,YYY,XZI,XIX,IZX\}, \eeq\noindent \beq
\{7,13,24,56,137,247,567\}
\leftrightarrow\{YXZ,YIZ,XXY,XIY,IXI,ZIX,ZXX\}, \eeq\noindent \beq
\{7,15,24,36,157,247,367\}
\leftrightarrow\{YXZ,XYZ,XXY,YYY,ZZI,ZIX,IZX\}, \eeq\noindent \beq
\{7,14,23,56,147,237,567\}
\leftrightarrow\{YXZ,YZX,XYI,XIY,IYY,ZZZ,ZXX\}, \eeq\noindent \beq
\{7,12,35,46,127,357,467\}
\leftrightarrow\{YXZ,ZYZ,ZYI,YXI,XZI,XZZ,IIZ\}, \eeq\noindent \beq
\{7,15,23,46,157,237,467\}
\leftrightarrow\{YXZ,XYZ,XYI,YXI,ZZI,ZZZ,IIZ\}, \eeq\noindent \beq
\{7,13,26,45,137,267,457\}
\leftrightarrow\{YXZ,YIZ,ZIY,ZXY,IXI,XXX,XIX\}. \label{lastorbit}
\eeq\noindent
Using the cyclic shift and the
corresponding action of $\mathcal{D}(\alpha)$, we can generate
seven more planes for each member of this set of $15$ planes.
Hence the number of Fano planes of this kind is $7\times 15=105$.

A Fano plane of the second class consists solely of symmetric operators, i.\,e. triples of numbers from $1$ to $7$. Since the automorphism group of the Fano plane
has order $168$, the number of such Steiner triples is $7!/168=30$.
The set of triples $124,235,346,457,561,672$ and $713$ corresponds to the
pairwise commuting set of Eq.\,(\ref{xesek}), featuring only $X$ and $I$, 
\beq \{124,235,346,457,561,672,713\}
\leftrightarrow\{XXI,IIX,IXX,XIX,XII,XXX,IXI\}.\label{SS1}\eeq
\noindent
Similarly, the set of triples $126,237,341,452,563,674$ and $715$ corresponds to the pairwise
commuting set of Eq.\,(\ref{zsek}), featuring only $Z$ and $I$,
\beq\{126,237,341,452,563,674,715\}\leftrightarrow\{ZII,ZZZ,IZI,IZZ,ZIZ,IIZ,ZZI\}.\label{SS2}
\eeq\noindent
Because these two particular planes are left invariant by $\mathcal{D}(\alpha)$,
no new planes can be generated from them. However, the
remaining $28$ planes of this class indeed arise from four distinguished ones
under the cyclic shift of $\mathcal{D}(\alpha)$. These distinguished planes are \beq
\{123,147,156,246,257,345,367\}\leftrightarrow\{XYY,IYY,XII,XZX,IXZ,XXZ,IZX\},\label{S1}\eeq\noindent
\beq
\{127,135,146,236,245,347,567\}\leftrightarrow\{XZI,ZYY,YXY,XIZ,IZZ,YYX,ZXX\},\label{S2}\eeq\noindent
\beq\{126,134,157,235,247,367,456\}\leftrightarrow\{ZII,IZI,ZZI,IIX,ZIX,IZX,ZZX\},\label{S3}
\eeq\noindent
\beq\{124,137,156,236,257,345,467\}\leftrightarrow\{XXI,IXI,XII,XIZ,IXZ,XXZ,IIZ\}\label{S4}.
\eeq\noindent
We have mentioned in the preceding section that the symmetric elements of $\mathcal{P}_3$ all lie on a particular hyperbolic quadric $\mathcal{Q}^{+}(5,2) \equiv \mathcal{Q}$ of the ambient projective space $PG(5,2)$.
We have also found that each plane of the second class features only symmetric elements; hence, all of them must by fully located on this particular quadric. Next, it is well known that 
the planes lying on any $\mathcal{Q}^{+}(5,2)$ split into two distinct systems, of cardinality 15 each.  Employing the famous Klein correspondence between the lines of $PG(3,2)$ and the points of the (Klein quadric) $\mathcal{Q}^{+}(5,2)$ \cite{Hirschfeld}, a plane from one system corresponds to the set of lines through a point of $PG(3,2)$, whereas a plane of the other system answers to the set of lines in a plane of $PG(3,2)$. From this correspondence it readily follows that two distinct planes belonging the same system have just a single point in common, whilst two planes of different systems are either disjoint, or share a line. Thus, our two special planes, Eq.\,(\ref{SS2}) and Eq.\,(\ref{SS1}), being disjoint, come from different systems. Further, the planes defined by Eqs.\,(\ref{SS2}), (\ref{S2}) and (\ref{S4}) are all from the same system, since their pairwise intersection is a single point. Clearly, this property is also exhibited by the remaining $12$ planes arising via a repeated action of the automorphism of order seven, which thus complete one system. Similarly, the $15$ planes defined by Eqs.(\ref{SS1}), (\ref{S1}) and
(\ref{S3}), together with their $12$ cyclically shifted cousins, belong all to the other system. For the reader's convenience, all the 30 planes lying on our Klein quadric $\mathcal{Q}$ will be explicitly listed later on (see Sect.\,6).

We conclude this section by noting that the action of $Sp(6,2)$ on $\mathcal{I}$ is transitive \cite{Cossking1}, that is, given any two planes from $\mathcal{I}$, one can find an element of $Sp(6,2)$ that sends one plane to the other.

\section{Planes, trivectors and the Grassmannian \qvec{Gr(6,\,3)}}

Having at our disposal a rather detailed description of the context space $\mathcal{I}$  and $Sp(6,2)$-action on it, we can now proceed to our second major task, namely the issue of mapping bijectively this space into the one of symmetric operators on four qubits. As already mentioned in the introduction, although the explicit form of this mapping has not yet been worked out, the geometric construction underlying it --- the so called spin module of the group $Sp(6,2)$  --- is well known in mathematical literature \cite{Gow,Cossking1,Cossidente}. In order to understand this construction, we will first provide another important representation of our context space. Obviously, the set of planes $\mathcal{I}$ can be regarded as a special subset of a total of 1395 planes living in $PG(5,2)$. So,  as a first step first, we will characterize this full set of planes in terms of $20$ Pl\"ucker coordinates, which are related to the independent components of separable trivectors.

The set of planes in $PG(5,2)$ comes from the projectivization of the set of three-dimensional subspaces in our six-dimensional vector space $V_3=\field{Z}_2^6$, i.\,e. from the Grassmannian $Gr(6,3)$; the projectivization of this latter space will be denoted by $\mathcal{G}r(5,2)$. Each element of $\mathcal{G}r(5,2)$ can be viewed
as the left row space of a $3\times 6$ matrix $(A\vert B)$ of
rank $3$, where $A$ and $B$ are $3\times 3$ matrices, whose entries are
taken from $\field{Z}_2$. The meaning of this term is as follows.
The three {\it rows} of the $3\times 6$ matrix $(A\vert B)$ 
can be regarded as the three linearly independent vectors spanning
a three-dimensional subspace in $V_3$, i.\,e. an element of
$Gr(6,3)$; equivalently, the corresponding points span a plane of $PG(5,2)$,
i.\,e. an element of $\mathcal{G}r(5,2)$. For example, the plane
$\{7,12,34,56,127,347,567\}$ of Eq.\,(\ref{C3}) is spanned by the
observables $12,34,56$, i.\,e. $ZYZ,IZY,XIY$. Their
corresponding vectors in $V_3$ are $(111010),(011001),(001101)$;
hence, we have 
\beq \{7,12,34,56,127,347,567\}\leftrightarrow
(A\vert B),\qquad
A=\begin{pmatrix}1&1&1\\0&1&1\\0&0&1\end{pmatrix},\quad
B=\begin{pmatrix}0&1&0\\0&0&1\\1&0&1\end{pmatrix}. \label{kodc3}
\eeq 
\noindent Under the {\it left action} via an element $T\in
GL(3,\field{Z}_2)$, we obtain a new $3\times 6$ matrix
$(A^{\prime}\vert B^{\prime})=T(A\vert B)$ that, obviously, represents the {\it same} plane of $\mathcal{G}r(5,2)$. We express
this property symbolically as $(A\vert B)\simeq (A^{\prime}\vert
B^{\prime})$. As $GL(3,\field{Z}_2)$ is isomorphic to the automorphism group of the Fano
plane, it merely permutes the basis elements of a given plane from $\mathcal{G}r(5,2)$.
For planes represented by a matrix $(A\vert B)$ such that $B\in
GL(3,\field{Z}_2)$, i.\,e. when $B$ is invertible, one can use the
matrix $(M\vert I_3)$  as a representative spanning the same
plane. Here, $M=B^{-1}A$ and $I_3$ is the $3\times 3$ identity
matrix. For example, the plane defined by Eq.\,(\ref{kodc3}) is characterized by ${\rm
Det}B=1$; hence, $B\in GL(3,\field{Z}_2)$ and a short calculation yields
\beq (A\vert B)\simeq (M,I_3),\qquad
M=B^{-1}A=\begin{pmatrix}0&1&0\\1&1&1\\0&1&1\end{pmatrix},\label{kodc32}
\eeq \noindent and the three rows of the new matrix $(M\vert I_3)$
thus define another triple of operators, namely $XZI,ZYZ,IZY$.
Since, according to Eq.\,(\ref{C3}), these operators show up in the list of seven
operators and they are not collinear, they span the same plane. On
the other hand, all planes with coordinates $(A\vert B)$, where
$B\notin GL(3,\field{Z}_2)$, will be called {\it planes at
infinity}. The plane represented by the matrix $(I_3\vert 0)$ will
be called the {\it distinguished plane} of $PG(5,2)$. This is the plane
defined by Eq.\,(\ref{SS2}), forming an
orbit of its own under $\mathcal{D}(\alpha)$. One can show that a
plane is at infinity precisely when it has nonzero intersection
with this distinguished plane.

In the next step, we shall embed $\mathcal{G}r(5,2)$ into the
space of trivectors $\bigwedge^3\field{Z}_2^6$ using the well-known Pl\"ucker
embedding. An arbitrary trivector can be expressed as
\beq P=\sum_{1\leq\mu<\nu<\rho\leq 6}P_{\mu\nu\rho}e_{\mu}\wedge
e_{\nu}\wedge e_{\rho}. \eeq \noindent Here, $P_{\mu\nu\rho}$ are
${6\choose 3}=20$ linearly independent expansion coefficients.
Since these $20$ numbers are in $\field{Z}_2$, the alternating
property now means symmetrization as well as vanishing of the
diagonal elements. Hence, the $20$ $P_{\mu\nu\rho}$s can be
extended to a rank three tensor whose indices are symmetric
under permutations, but which vanishes when any two indices happen to
be the same. An element $P\in \bigwedge^3\field{Z}_2^6$ is called
separable if it can be written in the form $P=u\wedge v\wedge w$
for some linearly independent elements $u,v,w\in V_3$. Hence, a
three-space of $Gr(6,3)$ corresponds to a separable three-form in
$\bigwedge^3\field{Z}_2^6$. Equivalently, a plane in
$\mathcal{G}r(5,2)$ corresponds to a point in the subset of
separable trivectors in the projectivization of
$\bigwedge^3\field{Z}_2^6$, which is a 19-dimensional projective space over $\field{Z}_2$, $PG(19,\,2)$.
Explicitly, the Pl\"ucker embedding $\theta$ is given by
the map \beq \theta: Gr(6,3)\hookrightarrow
\bigwedge^3\field{Z}_2^6,\qquad
(M\vert N)=\begin{pmatrix}u_1&u_2&u_3&u_4&u_5&u_6\\
v_1&v_2&v_3&v_4&v_5&v_6\\w_1&w_2&w_3&w_4&w_5&w_6\end{pmatrix}\hookrightarrow
u\wedge v\wedge w. \label{pluckerembedding} \eeq 
\noindent Using
the canonical basis vectors $e_{\mu}$ defined by Eq.\,(\ref{bazis}), we have 
\beq
u\wedge v\wedge w=P_{123}e_1\wedge e_2\wedge e_3+P_{124}e_1\wedge
e_2\wedge e_4+\dots +P_{456}e_4\wedge e_5\wedge e_6.
\label{explicitalak} \eeq \noindent Hence, for a separable
trivector, $P_{\mu\nu\rho}$ are the $3\times 3$ minors of the matrix
$(M\vert N)$ obtained by keeping merely the columns of this
$3\times 6$ matrix labelled by the fixed numbers $\mu,\nu,\rho$;
they are called the Pl\"ucker coordinates of the given plane.
Clearly, the Pl\"ucker coordinates are
not independent. They are subject to quadratic relations, called
the Pl\"ucker relations. It is known (see, for example, \cite{GKZ}) that an {\it arbitrary}
$P\in\bigwedge^3\field{Z}_2^6$ is separable if, and only if, its
coefficients $P_{\mu\nu\rho}$ satisfy the Pl\"ucker relations.
In our special case these relations can elegantly be described as
follows.
 
For an arbitrary $3\times 3$ matrix $M$, let us denote by
$M^{\sharp}$ the transposed cofactor matrix of $M$. Then, we have
\beq MM^{\sharp}=M^{\sharp}M={\rm Det}(M)I_3,\qquad
(M^{\sharp})^{\sharp}={\rm Det}(M)M,\qquad
(MN)^{\sharp}=N^{\sharp}M^{\sharp}, \label{relationsneeded} \eeq
\noindent 
and
\beq {\rm Det}(M+N)={\rm Det}M+{\rm
Tr}(MN^{\sharp})+{\rm Tr}(MN^{\sharp})+{\rm
Det}N.\label{relationsneeded2} \eeq\noindent
For a plane that can be represented in the form $(M\vert
I_3)$, the  Pl\"ucker coordinates can be conveniently arranged as two
numbers and two $3\times 3$ matrices as follows \beq P_{123}\equiv
{\rm Det}M,\quad
\begin{pmatrix}P_{156}&P_{256}&P_{356}\\P_{146}&P_{246}&P_{346}\\
P_{145}&P_{245}&P_{345}\end{pmatrix}=M,\quad
\begin{pmatrix}P_{234}&P_{235}&P_{236}\\P_{134}&P_{135}&P_{136}\\
P_{124}&P_{125}&P_{126}\end{pmatrix}=M^{\sharp},\quad P_{456}=1.
\label{szereposztas}\eeq \noindent Let us refer to these
quantities as the four-tuple \beq (m,M,N,n)\equiv ({\rm
Det}M,M,M^{\sharp},1).\label{Freudenthal}\eeq\noindent
 Now, in light of the identities given by
(\ref{relationsneeded}), we have \beq
mM=N^{\sharp},\qquad nN=M^{\sharp},\qquad mnI_3=MN.
\label{Pluckerrel} \eeq\noindent 
It can be shown that these
expressions, quadratic in  Pl\"ucker coordinates, are nothing but the
usual Pl\"ucker relations. We have thus shown that the $\theta$-images
of planes of the form $(M\vert I_3)$ obey the Pl\"ucker
relations. Conversely, it can be shown \cite{Leclerc} that if
an arbitrary trivector $P$, whose Pl\"ucker coordinates are arranged 
into a four-tuple $(m,M,N,n)$, where
\beq
m=P_{123},\quad
M=\begin{pmatrix}P_{156}&P_{256}&P_{356}\\P_{146}&P_{246}&P_{346}\\
P_{145}&P_{245}&P_{345}\end{pmatrix},\quad
N=\begin{pmatrix}P_{234}&P_{235}&P_{236}\\P_{134}&P_{135}&P_{136}\\
P_{124}&P_{125}&P_{126}\end{pmatrix},\quad n=P_{456},
\label{ujszerep} \eeq \noindent meets the constraints given by (\ref{Pluckerrel}), then $P$ is separable. Hence, Eq.\,(\ref{Pluckerrel}) can be used as a sufficient and necessary condition for the separability of a trivector. Such trivectors  can be identified with the
Grassmannian $Gr(6,3)$ via the  Pl\"ucker embedding $\theta$. From the projective viewpoint, the set of planes in $PG(5,2)$ is identified with the set of points of a certain algebraic
variety of $PG(19,2)$; this variety is defined by  Eq.\,(\ref{Pluckerrel}).

Further, one can define, in a complete analogy to what we did in the case of $PG(5,\,2)$, a symplectic polarity also on $PG(19,\,2)$. This
polarity originates from a symplectic form $\mathcal{B}$ defined on the associated $20$-dimensional vector space $\mathcal{V}_{10}$
over $\field{Z}_2$, 
\beq \mathcal{B}:\mathcal{V}_{10}\times
\mathcal{V}_{10}\to \field{Z}_2,\quad
((m,M,N,n),(m^{\prime},M^{\prime},N^{\prime},n^{\prime}))\mapsto
mn^{\prime}+nm^{\prime}+{\rm Tr}(MN^{\prime}+NM^{\prime}).
\label{Bforma} \eeq \noindent Here, the coordinates of a vector
of $\mathcal{V}_{10}$ are given in the form of (\ref{ujszerep}).
Moreover, in analogy to Eq.\,(\ref{kankvadrat}) we can also define a quadratic
form associated with $\mathcal{B}$, \beq q_0:
\mathcal{V}_{10}\to \field{Z}_2,\qquad (m,M,N,n)\mapsto mn+{\rm
Tr}(MN). \label{ujkvadratikus} \eeq \noindent Notice that, formally,
one can regard the vector space
$\mathcal{V}_{10}$  as the set of $10$-qubit
Pauli operators (defined up to a sign). Then, as usual, the symplectic
form $\mathcal{B}$ vanishes for commuting and differs from zero for
anticommuting pairs of such operators. Likewise, the quadratic form $q_0$
again gives zero for symmetric and one for antisymmetric
operators. When $(m,M,N,n)$ represents a
separable trivector, one can employ Eq.\,(\ref{Pluckerrel}) to see
that there exists a plane represented as $(A\vert B)$ such that
\beq m={\rm Det}A,\quad M=B^{\sharp}A,\quad N=A^{\sharp}B,\quad
n={\rm Det}B. \label{ABMN} \eeq \noindent In this special case
\beq q_0((m,M,N,n))={\rm Det}A{\rm Det}B+{\rm
Tr}(B^{\sharp}AA^{\sharp}B)= {\rm Det}A{\rm Det}B+
 {\rm Det}A{\rm Det}B{\rm Tr}I_3=0.
\label{symmetricfeltetel} \eeq \noindent This implies that the
planes of $PG(5,2)$ are mapped to those points of $PG(19,2)$ that
are lying on a certain hyperbolic quadric, viz. the one that accommodates all {\it symmetric} operators of $10$-qubit Pauli group.

It is obvious that $\mathcal{I}$, being a subset of planes of $PG(5,\,2)$, will be mapped by $\theta$ to a subvariety of the variety defined by Eq.\,(\ref{Pluckerrel}).
In order to find this subvariety, we make use of the symplectic polarity $\perp$ on $PG(5,2)$. Since the Pl\"ucker map sends planes to trivectors, planes of our context space will be represented by special trivectors. To see this, we first notice that  the action of $Sp(6,2)$
is no longer irreducible on the $20$-dimensional space of trivectors. It can be shown that the $20$-dimensional
representation of the group, induced by its particular representation on $V_3$, decomposes
as $20=6\oplus 14$. In order to properly grasp this decomposition, let us
introduce a bivector associated to the symplectic form $J$ in the following way \beq J=\sum_{1\leq\mu<\nu\leq 6}J_{\mu\nu}e_{\mu}\wedge
e_{\nu}, \label{bivector} \eeq \noindent where $J_{\mu\nu}$ is
given by Eq.\,(\ref{simplmatr}). Now, $\wedge^3V_3$ decomposes as
\beq \wedge^3V_3=J\wedge V_3+\wedge_0^3V_3,\qquad P_0\in
\wedge_0^3V_3\quad{\rm iff}\quad J\wedge P_0=0. \label{sdecomp}
\eeq \noindent The trivectors $P_0\in\wedge^3_0V_3$ are called
primitive and span the $14$-dimensional irreducible subspace.
Writing out the constraint $J\wedge P_0=0$ explicitly shows that,
in terms of the components $P_{\mu\nu\rho}$, the condition for
primitivity can be expressed as an extra condition on the matrices
$M$ and $N$ of Eq.\,(\ref{ujszerep}), namely \beq M^T=M,\qquad N^T=N;
\label{szimmcond} \eeq \noindent that is, these matrices become
{\it symmetric}.
A brute-force calculation shows that each plane from $\mathcal{I}$ is mapped by $\theta$  to
a primitive trivector satisfying Eq.\,(\ref{szimmcond}). Here is a quick
demonstration for the special case when either ${\rm Det}A$, or
${\rm Det }B$, is nonzero. One first  notes that for $(A\vert B)\in \mathcal{I}$
the row vectors are pairwise orthogonal; this implies that  $(A\vert B)J(A\vert
B)^T=0$ and so $AB^T=BA^T$. If ${\rm Det}B\neq 0$, we can use
an equivalent description of this plane as $(M\vert I_3)$, where
$M=B^{\sharp}A$. Now, $(M\vert I_3)\in \mathcal{I}$, so $M^T=M$
and because $N=M^{\sharp}$ (see Eq.\,(\ref{szereposztas})),  we
have $N^T=N$, too. For ${\rm Det}A \neq 0$ we use $(I_3\vert N)$
with $N=A^{\sharp}B$ to arrive at the same result.

The upshot of these considerations is as follows. Take a
particular plane of $\mathcal{G}(5,2)$, represented in the form
$(A\vert B)$. Calculate its Pl\"ucker coordinates, and
arrange them into a four-tuple $(m,M,N,n)$ using
Eq.\,(\ref{ujszerep}). If the plane belongs to $\mathcal{I}$, then
the corresponding matrices $M$ and $N$ will be symmetric. As a consequence,
only $14$ Pl\"ucker coordinates suffice to represent a plane from $\mathcal{I}$. This implies that the $\theta$-image of $\mathcal{I}$  spans in $PG(19,2)$ a projective subspace of
dimension $13$, $PG(13,\,2)$.

\section{Mapping the context space to symmetric four-qubit operators}

Because the vector space associated with the projective subspace
representing $\mathcal{I}$ is $14$-dimensional, we will refer to it as
$\mathcal{V}_7$.  It is instructive to calculate the restriction of
the symplectic form $\mathcal{B}$ (Eq.\,(\ref{Bforma})) to
$\mathcal{V}_7$. Let $\xi,\xi^{\prime}\in \mathcal{V}_7$, where $\xi\equiv(m,M,N,n)$,
$\xi^{\prime}=(m',M',N',n')$ and both  $M$ and $N$ are {\it symmetric}. By virtue of Eq.\,(\ref{ujszerep}), we get 
\begin{eqnarray}
\mathcal{B}(\xi,\xi^{\prime})&=&P_{123}P^{\prime}_{456}+
P_{456}P^{\prime}_{123}\\\nonumber
&+&P_{156}P^{\prime}_{234}+P_{234}P^{\prime}_{156}+P_{246}P^{\prime}_{135}
+P_{135}P^{\prime}_{246}+
P_{345}P^{\prime}_{126}+P_{126}P^{\prime}_{345};
 \end{eqnarray} \noindent
that is, because we are over $\field{Z}_2$ and all off-diagonal elements of
$M$ and $N$  occur in doubles, only the {\it diagonal elements} of these matrices are nonzero. This implies that whether two
elements of $\mathcal{V}_7$ are orthogonal or not is determined
merely by {\it eight} numbers comprising $m$, $n$ and the six diagonal
elements of $M$ and $N$. Rephrased in the language of Pauli
operators, the fact whether two seven-qubit operators commute or not
is determined solely by the relevant four-qubit part.

In order to isolate this important four-qubit part, let us split
our $\mathcal{V}_7$ into an $8$- and a
$6$-dimensional vector subspace as \beq \mathcal{V}_7=\mathcal{V}_4\oplus
\mathcal{V}_3. \label{splitting} \eeq \noindent Here, the elements of
$\mathcal{V}_4$ are of the form \beq
(P_{123},P_{156},P_{246},P_{345},P_{456},P_{234},P_{135},P_{126})
\label{ezekkellenek}\eeq \noindent and the elements of
$\mathcal{V}_3$ have the following representatives \beq
(P_{146},P_{245},P_{356},P_{235},P_{136},P_{124})=(P_{256},P_{346},P_{145},P_{134},P_{125},P_{236}).
\eeq\noindent 
Under this ordering of the components for the four-qubit
part, the restricted symplectic form
$\mathcal{B}$ features a matrix similar to Eq.\,(\ref{simplmatr}), where
$I_3$ is now replaced by the matrix $I_4$ in the
off-diagonal blocks.
Clearly, to an element of $\mathcal{V}_4$ one can associate a
four-qubit operator defined, up to a sign, as \beq
(P_{123},P_{156},P_{246},P_{345},P_{456},P_{234},P_{135},P_{126})=(a_1a_2a_3a_4b_1b_2b_3b_4).
\label{hozzarendeles} \eeq \noindent 
Here, the pair $(a_ib_i)$,
$i=1,2,3,4$, corresponds to the $j$-th qubit, with the corresponding
operator given by the dictionary furnished by Eq.\,(\ref{corr1}). It is also
important to realize that, according to
Eq.\,(\ref{symmetricfeltetel}), the $\mathcal{V}_4$-restriction of the quadratic form 
(\ref{ujkvadratikus})  shows that
the relevant four-qubit operators  associated to planes
from $\mathcal{I}$ are {\it all} symmetric. This means that we
can establish an explicit mapping from $\mathcal{I}$ to the points
of a hyperbolic quadric defined by the zero locus of the quadratic form $q_0$ restricted to $PG(7,2)$. Let us
denote this hyperbolic quadric  by
$\mathcal{Q}^+(7,2)$. And because we have $135$ planes in $\mathcal{I}$
and there are precisely $135$ points on the
hyperbolic quadric $\mathcal{Q}^+(7,2)$ (see, for example, \cite{Edge}), this mapping should be --- and indeed {\it is} --- a
bijection.

In order to establish an explicit form of this bijection, we will
proceed as follows. We take our list of planes defined by
Eqs.\,(42)--(62). For each of these $21$
representative planes, one picks up three non-collinear three-qubit
operators. They define three linearly independent {\it row
vectors}, which we arrange as in Eq.\,(\ref{pluckerembedding}). These
vectors then generate the $3\times 6$ matrix $(A\vert B)$. Subsequently, after calculating minors with {\it columns} being just
the labels of $P_{\mu\nu\rho}$, we determine the $8$ relevant
Pl\"ucker coordinates of Eq.\,(\ref{ezekkellenek}). Then, using the dictionary given by (\ref{hozzarendeles}), we read off the corresponding
four-qubit operator. The results of our calculation can be summarized as
\beq\{7,16,25,34,167,257,347\}\mapsto YYXZ, \label{F1}\eeq\noindent
\beq\{7,16,24,35,167,247,357\}\mapsto YIYX, \label{F2}\eeq\noindent
\beq\{7,12,34,56,127,347,567\}\mapsto YIZY, \label{F3}\eeq\noindent
\beq\{7,14,25,36,147,257,367\}\mapsto YYII, \eeq\noindent
\beq\{7,16,23,45,167,237,457\}\mapsto IYZY, \eeq\noindent
\beq\{7,13,25,46,137,257,467\}\mapsto IIXZ, \eeq\noindent
\beq\{7,15,26,34,157,267,347\}\mapsto IYYX, \eeq\noindent
\beq\{7,14,26,35,147,267,357\}\mapsto XXYY, \eeq\noindent
\beq\{7,12,45,36,127,457,367\}\mapsto XXZX, \eeq\noindent
\beq\{7,13,24,56,137,247,567\}\mapsto XZXI, \eeq\noindent
\beq\{7,15,24,36,157,247,367\}\mapsto ZZZX, \eeq\noindent
\beq\{7,14,23,56,147,237,567\}\mapsto ZZYY, \eeq\noindent
\beq\{7,12,35,46,127,357,467\}\mapsto ZXIZ, \eeq\noindent
\beq\{7,15,23,46,157,237,467\}\mapsto ZXXI, \eeq\noindent
\beq\{7,13,26,45,137,267,457\}\mapsto XZIZ, \eeq\noindent
\beq\{123,147,156,246,257,345,367\}\mapsto XXII, \eeq\noindent
\beq\{127,135,146,236,245,347,567\}\mapsto ZIZZ, \eeq\noindent
\beq\{126,134,157,235,247,367,456\}\mapsto IIIX, \eeq\noindent
\beq\{124,137,156,236,257,345,467\}\mapsto IIIZ, \eeq\noindent
\beq\{124,235,346,457,561,672,371\}\mapsto XIII, \eeq \noindent
\beq \{126,237,341,452,563,674,715\}\mapsto ZIII. \label{F21}\eeq
\noindent
Note that the last two planes are the special ones,
Eqs.\,(\ref{SS1})--(\ref{SS2}), fixed by the automorphism of
order seven. The $19$ remaining planes generate the remaining elements of $\mathcal{I}$ via this
automorphism. We know that this automorphism acts on the left-hand-side of the bijection as a cyclic shift of order
seven. In order to find the four-qubit labels of the
remaining planes, we need to figure out how this automorphism acts on the right-hand-side.

To this end in view, one observes that the $6\times 6$ matrix of
this automorphism  acts on a plane of the form $(A\vert B)$ as \beq (A\vert B)\mapsto
(A^{\prime}\vert B^{\prime})=(AK\vert BL),\label{igyhat} \eeq
\noindent where $K$ and $L$ are the matrices known from
Eq.\,(\ref{dalfa}). Notice that ${\rm Det}K={\rm Det}L=1$ and
$L^{\sharp}=K^T$. Then $m^{\prime}={\rm Det}A^{\prime}={\rm
Det}A{\rm Det}K=m$, and $n^{\prime}={\rm Det}B^{\prime}={\rm
Det}B{\rm Det}L=n$. Hence, the first and the fourth coordinate in Eq.\,(\ref{hozzarendeles}) does not change, 
\beq
P^{\prime}_{123}=P_{123},\qquad
P^{\prime}_{456}=P_{456}.\label{0traf}\eeq\noindent 
Moreover, by virtue of (\ref{ABMN}), we have \beq
M^{\prime}=(BL)^{\sharp}AK=L^{\sharp}(B^{\sharp}A)K=K^TMK,\quad
N^{\prime}=(AK)^{\sharp}BL=L^TML.\quad \label{ks} \eeq \noindent
From the last equations we can extract the following transformation rules for diagonal
elements \beq P^{\prime}_{156}=P_{156}+P_{345},\quad
P^{\prime}_{246}=P_{246}+P_{156}+P_{345},\quad
P^{\prime}_{345}=P_{156}+P_{246},\label{1traf} \eeq \noindent \beq
 P^{\prime}_{234}=P_{234}+P_{135},\quad
P^{\prime}_{135}=P_{135}+P_{234}+P_{126},\quad
P^{\prime}_{126}=P_{126}+P_{234}.\label{2traf} \eeq \noindent
Since we work over $\field{Z}_2$, matrices $M$ and $N$ do not exhibit any mixing of diagonal and off-diagonal entries. 
One can also describe these transformation rules by the
$8\times 8$ matrix \beq R(\alpha)=
\begin{pmatrix}1&0&0&0&0&0&0&0\\
               0&1&1&1&0&0&0&0\\
               0&0&1&1&0&0&0&0\\
               0&1&1&0&0&0&0&0\\
               0&0&0&0&1&0&0&0\\
               0&0&0&0&0&1&1&1\\
               0&0&0&0&0&1&1&0\\
               0&0&0&0&0&0&1&1\end{pmatrix},\label{Ralfa}
               \eeq
               \noindent
acting {\it from the right} on row vectors of 
(\ref{hozzarendeles}). As this matrix also contains
the matrices $K$ and $L$ of Eq.\,(\ref{dalfa}), the
transformation rules of the four-qubit symmetric operators under
the automorphism of order seven are rather simple. The first
(leftmost) qubit is left invariant, whereas the second, third and the fourth
operators, regarded altogether as a three-qubit one, are cyclically shifted
according to the pattern we already know from
Eqs.\,(\ref{clifforbit})--(\ref{zsek}). Thus, for example,  a
cyclic shift sends the plane (\ref{F1}) to 
\beq
\{1,27,36,45,127,136,145\}\mapsto YIIY,\eeq
\noindent where for the shift of the last three operators of the four-qubit one we took into account Eq.\,(\ref{clifforbit}). 
We have thus completed our task of labeling the elements of the context space, $\mathcal{I}$,
in terms of symmetric four-qubit operators.
A brief inspection of Eqs.\,(\ref{F1})--(\ref{F21}) shows that two planes overlap when the corresponding four-qubit operators are
commuting. This is in accordance with the
proposition \cite{Cossidente} that two points lying on $\mathcal{Q}^+(7,2)$ are
perpendicular with respect to the symplectic form given by
Eq.\,(\ref{Bforma}) if, and only if, the corresponding planes from
$\mathcal{I}$ have non-empty intersection. As we have already seen, there are two
possibilities for this: either the two planes share a
point, or a line. An example of the first case is furnished by the planes (\ref{F2}) and (\ref{F3}), the
second case can be illustrated by the planes  (\ref{F1}) and ({\ref{F2}). For
an example of two disjoint planes one can consider the planes (\ref{F1}) and (\ref{F21}), for the corresponding four-qubit
observables  labeling these planes are anticommuting. 

At this point we will make a slight digression from our main line of reasoning and consider a {\it spread of  planes} of $PG(5,\,2)$, that is, a set of pairwise disjoint planes partitioning its point-set. From the physical point of view, such a spread is a partition of the $63$ non-trivial observables of $\mathcal{P}_3$ into nine pairwise {\it disjoint} heptads.
As an illustrative example, we can take the following set
\beq \{ \{XZY,ZYY,YXI,YXY,ZYI,XZI,IIY\},
\nonumber\eeq\noindent \beq \{YII,IYY,YYY,IZX,YZX,YXZ,IXZ\},
\nonumber\eeq\noindent\beq \{ZXX,IXX,ZII,IYZ,ZZY,IZY,ZYZ\},
\nonumber\eeq\noindent\beq \{YXZ,IXI,ZIZ,YIX,XXY,XIY,YXX\},
\nonumber\eeq\noindent\beq \{ZIY,ZXI,IXY,YYX,XYZ,YZZ,XZX\}, \eeq
\noindent\beq \{XXI,YYI,ZZI,XXZ,YYZ,IIZ,ZZZ\},
\nonumber\eeq\noindent\beq \{XYY,ZYX,YIZ,IZZ,ZXY,YZI,XXX\},
\nonumber\eeq \noindent \beq \{ZZX,ZIX,IZI,XZZ,YIY,XIZ,YZY\},
\nonumber\eeq\noindent\beq \{IYX,XYI,IXX,XYX,XIX,XII,IYI\}\}.
\nonumber\eeq\noindent Using our dictionary,  we readily find that this spread corresponds to the following set of four-qubit observables
\beq
\{YXZY,YYII,IZXX,IZXZ,YZIY,IXXI,IXYY,ZIZI,XIZI\}.
\eeq\noindent 
A quick check shows that these observables $\gamma_r, r=1,2,\cdots, 9$, are pairwise
anticommuting and each squares to $I_{16}$; hence,
$\{{\gamma}_r,\gamma_s\}=2\delta_{rs}I_{16}$, i.\,e. they form
the basis vectors of a $Cliff(9)$. Geometrically speaking, they represent an {\it ovoid} of $\mathcal{Q}^{+}(7,2)$ (see, e.\,g., \cite{Edge,ovoid}).
It is known that there are
$960$ such spreads/ovoids, hence the number of possible basis
vectors for a $Cliff(9)$ algebra made entirely from symmetric
four-qubit observables is $960$ as well. We also mention that the notion of a
spread of planes of $PG(5,\,2)$ is, in the three-qubit case, intimately related to the very important notion of mutually
unbiased bases.

For the sake of completeness, we will also present the $8\times 8$ representation of the remaining generator of
$Sp(6,2)$, $\beta$, which is of order two. In order to
calculate the relevant matrix, we rewrite the $6\times 6$ symplectic
matrix given by Eq.\,(\ref{betamatrix}) in a block form consisting of
$3\times 3$ matrices $a,b,c,d$ as \beq
D(\beta)=\begin{pmatrix}a&b\\c&d\end{pmatrix},
\eeq \noindent where the individual blocks can be readily read off from Eq.\,(\ref{betamatrix}). Then,
under the transformation $(A\vert B)\mapsto (Aa+Bc\vert Ab+Bd)$, we
get
\begin{eqnarray}
 m^{\prime}&=&{\rm Det}(Aa+Bc)={\rm Det}(Aa)+{\rm
Tr}(Aac^{\sharp}B^{\sharp})+{\rm Tr}(Bca^{\sharp}A^{\sharp})+{\rm
Det}(Bc)\\&=&{\rm Det}A+{\rm Tr}(ca^{\sharp}A^{\sharp}B)=m+{\rm
Tr}(ca^{\sharp}N),\end{eqnarray} \noindent
where we have taken into account that, according to Eq.\,(\ref{betamatrix}), ${\rm
Det}a=1$, ${\rm Det}c=0$, and $c^{\sharp}=0$, and also employed definition (\ref{ABMN}). Using the explicit forms of the matrices $a$ and
$c$ as well as expressions (\ref{ujszerep}), one finds \beq
\tilde{P}_{123}=P_{123}+P_{234}+P_{135}.\label{beta1} \eeq
\noindent Similar manipulations yield the transformation laws \beq
\tilde{P}_{456}=P_{456}+P_{345}, \label{beta2}\eeq\noindent \beq
\tilde{P}_{156}=P_{156}+P_{135}+P_{456}, \quad
\tilde{P}_{246}=P_{246}+P_{234}+P_{456},\quad
\tilde{P}_{345}=P_{345} \label{beta3},\eeq\noindent
 \beq
\tilde{P}_{234}=P_{234}+P_{345}, \quad
\tilde{P}_{135}=P_{135}+P_{345},\quad
\tilde{P}_{126}=P_{126}+P_{246}+P_{156}+P_{123};
\label{beta4}\eeq\noindent hence, the corresponding matrix that acts
on row vectors from the right has the form \beq R(\beta)=
\begin{pmatrix}1&0&0&0&0&0&0&1\\
               0&1&0&0&0&0&0&1\\
               0&0&1&0&0&0&0&1\\
               0&0&0&1&1&1&1&0\\
               0&1&1&0&1&0&0&0\\
               1&0&1&0&0&1&0&0\\
               1&1&0&0&0&0&1&0\\
               0&0&0&0&0&0&0&1\end{pmatrix}\label{Rbeta}.
               \eeq
               \noindent
One can verify that $R^2(\beta)=I_8$.

What we have constructed here is an explicit realization of the so-called spin module, or spin representation, of
$Sp(6,2)$ \cite{Cossking1,Gow}. In the mathematical literature, the $8$-dimensional
representation space for $Sp(6,2)$ that corresponds to our
$\mathcal{V}_4$ is constructed as a quotient space of the $14$-dimensional space $\mathcal{V}_7$ with respect to the unique
maximal subspace $\mathcal{V}_3$, fixed by $Sp(6,2)$.

Finally, it is also worth noticing that the block-diagonal nature of $R(\alpha)$ corresponds
to the fact that $D(\alpha)$ of Eq.\,(\ref{dalfa}) gives a
representation for one of the generators of $GL(3,2)=SL(3,2)$, the latter being
a subgroup of $Sp(6,2)$ consisting of block-diagonal
$6\times 6$ matrices. Then using instead  $K$ and
$L$ of Eq.\,(\ref{dalfa}) any two matrices, say $a$ and $d$, of
$GL(3,2)$, related as $a^{\sharp}=d^T$, the corresponding action of
an element of $SL(3,2)$ on the four-qubit operators is just the
usual action coming from the one that can be constructed on
three-qubits. This means that the first entry of a four-qubit
operator is left invariant, and the last three ones are
transformed according to this particular three-qubit representation. Such a
construction then trivially leads to an $SL(3,2)$-representation on
the four-qubit counterparts of the elements of $\mathcal{I}$.

\section{Mermin's pentagrams}

Our formalism has now been developed to such an extent that it can be employed to gain fundamental insights into the structure of so-called Mermin's pentagrams, objects living inside our
symplectic polar space $\mathcal{W}(5,2)$ and central to the conceptual issues related to quantum contextuality. Introduced by Mermin \cite{Mermin1}, a Mermin's pentagram is a configuration
consisting of ten three-qubit operators arranged along five edges sharing pairwise a single point. Each edge 
features four operators that are pairwise commuting and whose product is $+III$ or $-III$, with the understanding that the latter possibility occurs an odd number of times. 
A recent computer search \cite{Holweck} has shown that $\mathcal{W}(5,2)$ contains altogether $12096$ Mermin's pentagrams, of which $336$ are formed of solely {\it symmetric} observables. 
It was also pointed out that these numbers are rather remarkable, since $12096$ is the
order of the group $G_2(2)$, which is the automorphism group of the smallest
split Cayley hexagon, and $336$ is just the twice of the order of $SL(3,2)$, the latter being the automorphism group of the smallest projective plane.
We shall, among other things, provide an elegant computer-free justification of the occurrence of the second number.

To begin with, one recalls \cite{SanigaLevay}  that an edge of
a pentagram represents an affine plane of order two, i.\,e. the plane that originates from the Fano plane via omitting one
of its lines. Now, as each Fano plane gives birth to seven such affine planes and $\mathcal{I}$ features $135$ Fano planes, then we have altogether
$945$ copies of affine planes, each a possible candidate for an edge of a
Mermin's pentagram. In this pool of affine planes we will look for such quintuples that have the above-described intersection property; every such quintuples will thus be a potential candidate for a Mermin's pentagram.

To this end in view, we will first have a look at the set of $30$ planes that are lying on our particular Klein quadric $\mathcal{Q}$, accommodating all symmetric three-qubit observables.
As already described (see Sect.\,3), these planes form two distinct systems of cardinality 15 each. One system (let us call it $\mathcal{L}$) consists of
\beq
\{XII,XYY,IYY,XZX,IXZ,XXZ,IZX\}\leftrightarrow XXII,\label{K1}\eeq
\noindent
\beq
\{XXX,ZXZ,YIY,XZZ,YYI,ZZX,IYY\}\leftrightarrow XXXX,\label{K2}\eeq
\noindent
\beq
\{IXI,XXZ,XIZ,YXY,ZIX,ZXX,YIY\}\leftrightarrow XIXI,\label{K3}\eeq
\noindent
\beq
\{XXI,ZZX,YYX,IXZ,ZYY,YZY,XIZ\}\leftrightarrow XXXI,\label{K4}\eeq
\noindent
\beq
\{IIX,ZXX,ZXI,YYI,XZX,XZI,YYX\}\leftrightarrow XIIX,\label{K5}\eeq
\noindent
\beq
\{IXX,YZY,YYZ,ZIX,XZZ,XYY,ZXI\}\leftrightarrow XIXX,\label{K6}\eeq
\noindent
\beq
\{XIX,XZI,IZX,ZYY,YXY,ZXZ,YYZ\}\leftrightarrow XXIX,\label{K7}\eeq
\noindent
\beq
\{IIX,ZII,IZI,ZZI,ZIX,IZX,ZZX\}\leftrightarrow IIIX,\label{L1}\eeq
\noindent
\beq
\{IXX,ZZZ,IZZ,ZII,ZYY,IYY,ZXX\}\leftrightarrow IIXX,\label{L2}\eeq
\noindent
\beq
\{XIX,IZI,ZIZ,ZZZ,XZX,YIY,YZY\}\leftrightarrow IXIX,\label{L3}\eeq
\noindent
\beq
\{XII,IZZ,IIZ,IZI,XZZ,XIZ,XZI\}\leftrightarrow IXII,\label{L4}\eeq
\noindent
\beq
\{XXX,ZIZ,ZZI,IZZ,YXY,YYX,XYY\}\leftrightarrow IXXX,\label{L5}\eeq
\noindent
\beq
\{IXI,IIZ,ZII,ZIZ,IXZ,ZXI,ZXZ\}\leftrightarrow IIXI,\label{L6}\eeq
\noindent
\beq
\{XXI,ZZI,ZZZ,IIZ,YYI,YYZ,XXZ\}\leftrightarrow IXXI,\label{L7}\eeq
\noindent
\beq
\{XXI,IIX,IXX,XIX,XII,XXX,IXI\}\leftrightarrow XIII.\label{M}\eeq
\noindent
The $15$ planes of the other system (called $\mathcal{G}$) are those that feature swapped entries $Z$ and $X$ in labeling of both three- and four-qubit operators. 
Notice that for each plane from (\ref{K1})--(\ref{L7}) the last three entries of the four-qubit label are identical with the three-qubit label of the first element. One can see that all the 15 planes  share pairwise  a single point and the corresponding four-qubit operators are pairwise commuting. 
It is worth pointing out here that the special plane \beq \{ZII,ZZZ,IZI,IZZ,ZIZ,IIZ,ZZI\}\leftrightarrow ZIII\label{Z}\eeq\noindent is disjoint from the planes defined by Eqs.\,(\ref{K1})--(\ref{K7}) and Eq.\,(\ref{M}), since the corresponding four-qubit operators anticommute, and sharing a line with each of the planes (\ref{L1})--(\ref{L7}), where the corresponding four-qubit representatives commute.

Let us now consider pentads of pairwise commuting four-qubit operators from  $\mathcal{L}$ such that their product is $IIII$.
A handy example is the set $\{XXXX,XIII,IXII,IIXI,IIIX\}$. One can readily see that the ten three-qubit operators coming from pairwise intersections of the corresponding planes form
a pentagram, 
\beq
\{XXX,ZZX,ZXZ,XZZ\},\quad\{XXX,IIX,XII,IXI\},\quad\{XZZ,IIZ,XII,IZI\},\nonumber
\eeq
\noindent
\beq
\{ZII,ZZX,IIX,IZI\},\quad \{ZXZ,IIZ,ZII,IXX\}.\label{kanonikusp}
\eeq
\noindent
Another illustrative example is
$\{XIII,XXXX,IXXI,IIXI,IIXX\}$, whose associated pentagram looks as follows
\beq \{XXX,YYI,IYY,ZXZ\},\quad
\{XXX,XXI,IXI,IXX\},\quad\{ZXZ,ZII,IXI,IIZ\}, \eeq \noindent \beq
\{ZZZ,YYI,XXI,IIZ\},\quad\{ZZZ,IYY,ZII,IXX\}. \eeq \noindent
Notice that in both examples the four-qubit representatives of the
planes giving birth to pentagrams satisfy not only the
property $ABCDE=IIII$, but also the constraint that no three of
them are on a common line. Also, as the attentive reader might have noticed, in both examples the four-qubit observables feature only two different entries, namely $X$ and $I$.
Our next task will be to find all the other cases of this type. 

To this end, we express four-qubit labels in the form $X\otimes\mathcal{A}_i$ and
$I\otimes\mathcal{A}_i$, where $\mathcal{A}_i, i=1,2,\cdots, 7$, are
three-qubit operators that will be used to label the points of a
Fano plane. The remaining operator $X\otimes III$ is taken to have a special footing. 
Let us first focus on such quadruples of operators
$\{\mathcal{A}_1,\mathcal{A}_2,\mathcal{A}_3,\mathcal{A}_4\}$
that correspond to anti-flags of the Fano plane. An anti-flag consists of a line and a point not incident with that line; for example, the set $\{XXX,IIX,IXI,IXX\}$, where the point is
represented by $XXX$ and the line by the triple $\{IIX,IXI,IXX\}$. 
There are $28$ anti-flags in the Fano plane, each generating four pentagrams; hence, altogether $112$ pentagrams of this kind. The four pentagrams coming from the above-given example are 
\beq I\otimes XXX,\quad X\otimes XXX,\quad
X\otimes IXI,\quad X\otimes IIX,\quad X\otimes
IXX,\label{antiflag1} \eeq \noindent \beq I\otimes XXX,\quad
X\otimes XXX,\quad X\otimes IXI,\quad I\otimes IIX,\quad I\otimes
IXX,\label{antiflag2} \eeq \noindent \beq I\otimes XXX,\quad
X\otimes XXX,\quad I\otimes IXI,\quad X\otimes IIX,\quad I\otimes
IXX,\label{antiflag3} \eeq \noindent \beq I\otimes XXX,\quad
X\otimes XXX,\quad I\otimes IXI,\quad I\otimes IIX,\quad X\otimes
IXX\label{antiflag4}. \eeq \noindent 
The next kind of a quadruple $\{\mathcal{A}_1,\mathcal{A}_2,\mathcal{A}_3,\mathcal{A}_4\}$ corresponds to the complement of a line of the Fano plane, i.\,e. to the point-set of the associated affine plane of order two, which we will refer to as a quadrangle. We have seven such quadrangles. An example is the set $\{XXX,XII,IXI,IIX\}$, which is the complement
of the line $\{XXI,XIX,IXX\}$. Obviously, this construction yields two classes of
such pentagrams, and of cardinality $28$ each, which amounts to $56$ pentagrams of this kind.
For our particular example, the four-qubit labels of these
$4+4$ pentagrams are as follows \beq I\otimes XXX,\quad X\otimes
XII,\quad X\otimes IIX,\quad X\otimes IXI,\quad     X\otimes
III,\label{quad1} \eeq \noindent \beq X\otimes XXX,\quad I\otimes
XII,\quad I\otimes IXI,\quad I\otimes IIX,\quad X\otimes
III\label{quad2}, \eeq \noindent where the missing three
pentagrams from each class arise via a cyclic shift of
the operators $I$ and $X$ in the leftmost qubit to the remaining
members of the three-qubit operators belonging to the quadrangle.
Thus, for example, the next member of the class given by Eq.\,(\ref{quad1}) is $\{XXXX,IXII,XIIX,XIXI,XIII\}$. 
We thus arrive at the total of $168$ pentagrams coming from the planes of system $\mathcal{L}$. Following the same procedure with $X$ replaced by $Z$,
that is with the planes from the other system, $\mathcal{G}$, results in another set of $168$ pentagrams.
All in all, we find $336$ pentagrams that can be formed from symmetric
three-qubit observables. It represents no difficulty to verify that all these pentagrams are, in fact, Mermin's pentagrams. 
This is one of the major results found in \cite{Holweck} with the aid of a computer. Here, we have not only succeeded in furnishing a rigorous, computer-free explanation of this finding, but also shown that the whole set of ``symmetric'' pentagrams can be generated from merely {\it six} basic types, given by Eqs.\,(\ref{antiflag1})--(\ref{quad2}). 
Nay, noticing that the leftmost qubit must be associated with either two or four $X$
operators,  this classification can further be reduced to just {\it two}
kinds: namely,  anti-flag and quadrangle ones. Notice in passing that the special planes are tied uniquely to
the quadrangle kind.

An alternative explanation for the number $336$ goes as follows. We have seen that all the planes needed for construction of these
pentagrams lie on the Klein quadric, $\mathcal{Q}$, defined as the
zero locus $Q_0(v)=\sum_i^3a_ib_i=0$, where
$v=(a_1a_2a_3b_1b_2b_3)\in \field{Z}_2^6$. Under a transformation
of $SL(3,2)$ of the form (\ref{dalfa}), with $K$
and $L$ being replaced by nonsingular matrices $A$ and $D$ related to each other as 
$A^{\sharp}=D^T$, \beq Q_0={\bf a}{\bf b}^T\mapsto {\bf a}A({\bf
b}D)^T={\bf a}AD^T{\bf b}^T= {\bf a}{\bf b}^T, \eeq \noindent 
which means that $SL(3,2)$ leaves the Klein quadric invariant. On the other hand, the
transformation swapping the systems $\mathcal{L}$ and
$\mathcal{G}$ is the one with its $6\times 6$ matrix
representative being just the matrix $J$ of Eq.\,(\ref{simplmatr}).
This transformation also leaves $Q_0$, and so the
Klein quadric, invariant. One can actually prove that the group $SL(3,2)\cdot
2$ just described is a maximal subgroup of $G_2(2)$ \cite{Coop}. Moreover,
$SL(3,2)\cdot 2$ also lies inside the orthogonal group of
$\mathcal{Q}$, which is isomorphic to  $O^+(6,2)$. The facts that $SL(3,2)\cdot
2$ is  maximal inside $G_2(2)$
and that $O^+(6,2)$ possesses no subgroup isomorphic to $G_2(2)$ mean
that $SL(3,2)\cdot 2$ is the {\it full stabilizer} of $\mathcal{Q}$ in $G_2(2)$. 
Now,  $SL(3,2)$ is just the
stabilizer of the special planes (\ref{M}) and $(\ref{Z})$ that we
used in our construction of the $336$ pentagrams. The group
$SL(3,2)\cdot 2$, of order $336$, then acts transitively on these
planes by simply exchanging them. It can be shown \cite{Cossidente}
that this group also acts transitively on the set of planes
(\ref{L1})--(\ref{L7}), and separately on the set of planes
(\ref{K1})--(\ref{K7}). These facts strongly indicate that it
should be possible to use the $2\times 168=336$ elements of
$SL(3,2)\cdot 2$ to generate all the $336$ pentagrams from the
canonical one given by Eq.\,(\ref{kanonikusp}) and relate this group-theoretical method to the above-described geometric construction.

Obviously, the remaining pentagrams, which feature also antisymmetric
operators, can be generated by the repeated action of $Sp(6,2)$. For example, one can act on the canonical pentagram, Eq.\,(\ref{kanonikusp}), by $D(\beta)$ of
Eq.\,(\ref{betamatrix}) and on the corresponding four-qubit operators
by Eq.\,(\ref{Rbeta}), to obtain the pentagram \beq
\{ZII,IZI,IIX,ZZX\},\quad\{ZYX, ZZY,
ZXZ,ZII\},\quad\{YZX,ZZY,IZI,XZZ\},\nonumber\eeq \noindent \beq
\{XXX,ZXZ,XZZ,ZZX\},\quad\{ZYX,YZX,IIX,XXX\}.\eeq\noindent 
The four-qubit operators labeling the planes whose intersections yield this pentagram are $\{IIIX,ZZXI,ZXZI,XXXX,XZZI\}$. They again satisfy the identity $ABCDE=IIII$ and no three
of them are collinear. The five observables
are, of course, symmetric and pairwise commuting. By using the
action of $D(\alpha)$ and $D(\beta)$ of $Sp(6,2)$,
one can then generate new pentagrams. Alternatively, one can
generate the same pentagrams via the corresponding action of the
generators $R(\alpha)$ and $R(\beta)$ on the associated four-qubit operators.
Notice, however, that since the four-qubit operators are symmetric,
the spin representation $R$ of $Sp(6,2)$  on these operators can
be expressed as a conjugate action of type
(\ref{conjugateaction}). Accordingly, the condition
$ABCDE=IIII$ is preserved and, due to the symplectic nature
of these transformations, the remaining constraints on
the five four-qubit observables are left intact as well.

Can our approach also account for the total number of Mermin's pentagrams amounting to $12096$, i.\,e. the order of $G_2(2)$?
The authors of \cite{Holweck} made an intriguing conjecture that this number should stem form the properties of a remarkable point-line incidence geometry called the split Cayley hexagon of order two \cite{schr,psm}, which has $G_2(2)$ as its automorphism group.
Here, we can merely offer some remarks and conjectures on this issue, the details of which we would like to postpone to a separate paper.
First of all, notice that $G_2(2)$ is a maximal subgroup of $Sp(6,2)$. A  useful presentation for this group convenient for
our purposes is \cite{Vrana} \beq
G_2(2)=\langle\alpha,\gamma\rangle,
\qquad\gamma=\beta\alpha^2\beta\alpha\beta\alpha^3\beta\alpha^4\beta, \label{g2abr}\eeq \noindent 
where $\beta$ corresponds to the
transvection whose representative is given by $D(\beta)$ of
Eq.\,(\ref{betamatrix}) and $\alpha$ is the usual cyclic shift
generating the automorphism of order seven. Then,\footnote{We
thank Zsolt Szabo for checking these matrices for us on a
computer.} \beq
D(\gamma)=\begin{pmatrix}0&0&0&0&1&0\\
                         0&0&1&1&1&0\\
             1&0&0&1&0&1\\
             1&1&0&1&0&0\\
             1&0&1&0&1&0\\
             0&0&1&0&0&0\end{pmatrix},\qquad
R(\gamma)=\begin{pmatrix}0&0&0&1&1&0&1&0\\
                         0&0&0&0&0&0&1&0\\
             0&0&0&1&0&1&1&0\\
             1&1&0&0&1&1&0&1\\
                     1&0&0&1&0&0&1&0\\
                     1&1&1&0&1&1&0&0\\
             0&1&0&1&0&0&1&0\\
               0&0&0&1&0&0&0&0\end{pmatrix}.
             \label{dgamma}\eeq\noindent
These matrices are of order six. The matrix $R(\gamma)$ leaves invariant the special antisymmetric four-qubit operator $YIII$. Since, according to Eq.\,(\ref{Ralfa}), the other generator $R(\alpha)$ operates exclusively on the last three qubits via a cyclic shift and leaves $III$ invariant, $R(\alpha)$
and $R(\gamma)$ generate a maximal $G_2(2)$ subgroup of $Sp(6,2)$, leaving $YIII$ invariant.
This conforms to a theorem \cite{van}
that states that if we have a point lying off the $\mathcal{Q}^+(7,2)$, then its stabilizer within $Sp(6,2)$  is isomorphic to $G_2(2)$, and there is a single conjugacy class of $G_2(2)$'s in $Sp(6,2)$.
In our language of four-qubit observables this means that the stabilizer of each antisymmetric operator (which is not an element of our quadric accommodating only symmetric ones) defines a $G_2(2)$ subgroup. Since we have $120$  antisymmetric four-qubit operators,
there are $120$ possibilities for obtaining a $G_2(2)$ subgroup of $Sp(6,2)$.
This, clearly, reflects the fact that $\vert Sp(6,2)\vert/\vert G_2(2)\vert
=1451520/12096=120$.

\begin{figure}[t]
\centerline{\includegraphics[width=12truecm,clip=]{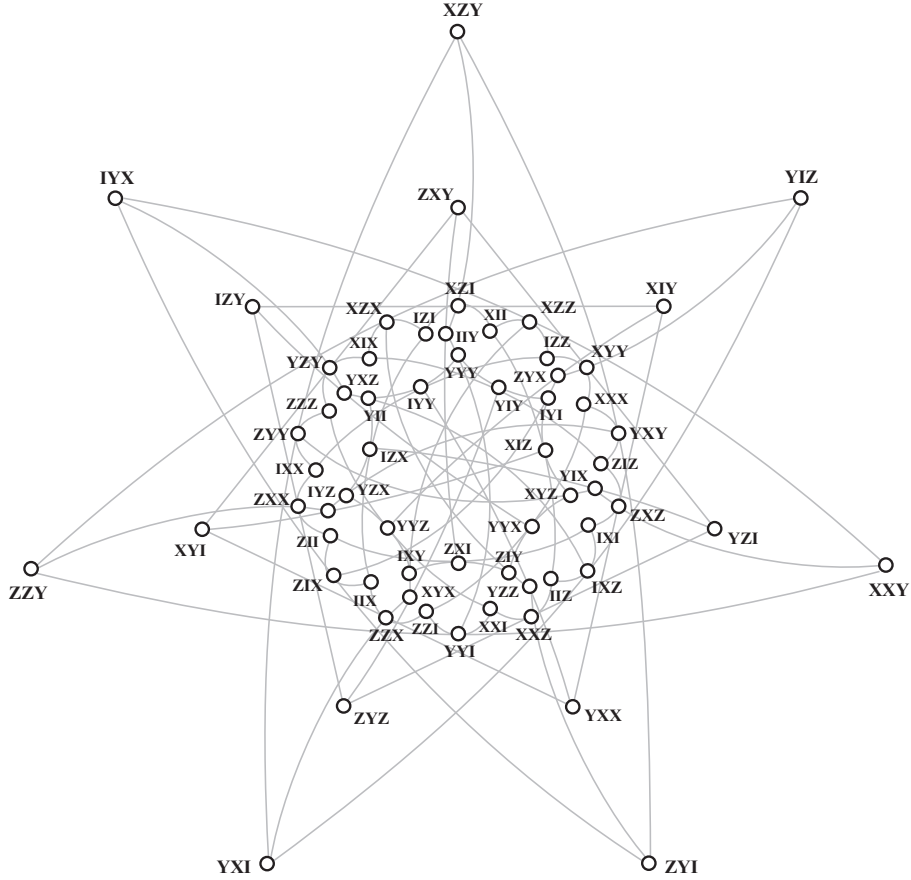}}
\caption{A diagrammatic illustration of the structure of the split Cayley hexagon of order two (based on drawings given in \cite{schr,psm}). The points are illustrated by small circles and its lines by triples of points lying on the same segments of straight-lines and/or arcs. Labeling by the elements of  $\mathcal{P}_3$ is adopted from \cite{LSV}. Also obvious is an automorphism of order seven of the structure.}
\end{figure}

A point of the four-qubit symplectic polar space, $\mathcal{W}(7,2)$, is collinear with 126 other points of this space (see, for example, \cite{Saniga1} and references therein).
If this point lies off the quadric $\mathcal{Q}^+(7,2)$, then 63 of these points will be located on the quadric itself, being at the same time the points of a copy of the split Cayley hexagon of order two. Let us now consider a particular set of 63 operators, each of which commutes with the special (antisymmetric) operator $YIII$: 
\beq Y\otimes
\mathcal{A},\quad \mathcal{A}^T=-\mathcal{A},\qquad I\otimes
\mathcal{S},\quad \mathcal{S}^T=\mathcal{S}. \label{hexapoints}
\eeq \noindent Here, $\mathcal{A}$ represents the set of $28$ 
antisymmetric and $\mathcal{S}$ stands for the set of $35$ nontrivial symmetric
three-qubit operators. Now, if one disregards the first-qubit labels, these 63 operators can be viewed as the 63 non-trivial elements of the three-qubit Pauli group, 
which were employed by two of us \cite{LSV} to label a copy of the split Cayley hexagon of order two when embedded in the corresponding three-qubit polar space $\mathcal{W}(5,2)$.
A diagrammatic illustration of the structure of our hexagon, together with the corresponding labeling, is shown in Figure 1.
Hence, as a representative of the hexagon living inside
$\mathcal{Q}^+(7,2)$ one can use the pictorial representation of Figure 1, with the only difference being that, according to
(\ref{hexapoints}), one has to also include the extra labels $Y$
and $I$ of the extra qubit. Then the $G_2(2)$ action on the points
and lines of this particular representation of the split Cayley hexagon of order two is generated by $R(\alpha)$ and $R(\gamma)$.

Having at our disposal an explicit form of the bijective correspondence between the points of
$\mathcal{Q}^+(7,2)$ and the planes of $\mathcal{I}$ (Sect.\,5), one can address the following interesting question: What kind
of triples of planes are the lines of the hexagon mapped to?
According to Theorem 3.4 of \cite{Cossidente}, the lines of
$\mathcal{Q}^+(7,2)$ are of two types; a line of one type arises from a pencil
of planes of $\mathcal{I}$, whereas that of other type comes from a plane-star on a fixed quadric. 
For example, a plane-star on our particular Klein quadric is any set of three planes in the same system ($\mathcal{L}$ or $\mathcal{G}$) that share a point, that is,
any triple of planes from Eqs.\,(\ref{K1})--(\ref{M}) such that when
their representative four-qubit operators are multiplied, the result is $IIII$. As an
example, one can take the planes defined by Eqs.\,(\ref{K1}), (\ref{K2}) and (\ref{L2}), labeled by  $XXII, XXXX$ and
$IIXX$, whose common point corresponds to $IYY$.  A line arising from a pencil of planes is, for example, the $\{IXIX,IIZI,IXZX\}$ one, as its
corresponding planes  \beq
\{XIX,IZI,ZIZ,ZZZ,XZX,YIY,YZY\},\nonumber\eeq\noindent \beq
\{IZI,IIX,XII,XIX,IZX,XZI,XZX\},\eeq\noindent
\beq\{YZZ,ZZY,ZIY,YIZ,XIX,XZX,IZI\},\nonumber\eeq\noindent 
share indeed a line, namely the  $\{XIX,XZX,IZI\}$ one.\footnote{It is interesting to note that the coordinates of this line are exactly those of the associated four-qubit operators with the first-qubit entry omitted.} A closer look at Figure 1 reveals that this line also belongs  to our hexagon.
Employing the formalism of \cite{LSV} and Theorem 4.1 of \cite{Cossidente}, it can be verified that {\it all} the lines of our split Cayley hexagon are of this ``pencil-of-planes'' type.

So, then, how is the aggregate of Mermin's pentagrams related to the split Cayley
hexagon of order two and its automorphism group $G_2(2)$? Clearly, the above considerations imply that out of the five
planes generating a Mermin's pentagram, no three can be in a
pencil of planes that corresponds to a line of the hexagon.
Moreover, as our hexagon picks up only $63$ planes from  $\mathcal{I}$, these particular
planes should somehow be used as a core set for labeling the
totality of pentagrams with  elements of $G_2(2)$. 
An investigation along these lines is under way and will be the subject of a separate paper.

\section{A link with the Black-Hole--Qubit Correspondence}

String/$M$-theory is the theory of extended objects, membranes and
strings. As it is well known, dynamics of such objects can
consistently be described provided that the ambient space-time has extra-dimensions. There exist different types of consistent
string theories, connected to each other by symmetries, called duality
symmetries \cite{Becker}. In the low-energy limit, these string
theories give rise to effective low-energy supersymmetric field
theories. When compactifying the low-energy effective actions, these
extra-dimensions are curled up into tiny compact spaces, and one is
left with the usual four-dimensional ``macroscopic''  space-time. 
Under the process of
curling up of the extra-dimensions, the wrapping configurations of
extended objects on nontrivial submanifolds of the compact space
manifest themselves via the occurrence of charges, of both magnetic and
electric type. There are also special scalar fields originating
from this mechanism, called moduli fields. They
come from fields describing the volume and shape of the
extra dimensions. The charges and moduli might form special
configurations that can give rise to special space-time curvature
effects, yielding charged extremal black holes in four dimensions.
There can be both supersymmetric and non-supersymmetric black
holes. In the case of toroidal compactifications, when the compact extra-dimensions are tiny tori of six dimensions for string- and
seven dimensions for $M$-theory, the resulting four-dimensional
theory is called $N=8$ supergravity.
 
It is also a well-known fact that the
most general class of charged, extremal black-hole solutions in
$N=8$ supergravity/$M$-theory in four dimensions is characterized by $56$
charges \cite{Becker}, equally-splitted into electric and
magnetic ones. These black-hole solutions are the ones of the
classical equations of motion of $N=8$ supergravity exhibiting an $E_{7(7)}$ symmetry, where $E_{7(7)}$ is the
non-compact real form of the exceptional group $E_7$ with the $56$
charges transforming according to its fundamental irreducible
representation. 
The corresponding black-hole solutions also display this
symmetry via their semiclassical Bekenstein-Hawking entropy
formulas, which are quartic polynomials invariant under $E_{7(7)}$.
At the level of quantum theory, the charges will be quantized and
the symmetry group will be the discrete subgroup $E_7(\field{Z})$,
called the $U$-duality group. An important subgroup of this group is $W(E_7)$. 
This Weyl group can be regarded as
the generalization of the usual group of electric -magnetic
duality, known from classical electrodynamics \cite{Pioline}.

As already stressed, $W(E_7)=Sp(6,2)/\field{Z}_2$ and since $Sp(6,2)$ has been shown to be intimately related to three-qubit observables, 
one may suspect that
the structure of the black-hole entropy and the $56$-dimensional
fundamental representation of $E_7$ can both be given a three-qubit-based reinterpretation. 
This is indeed the case. The relevant
reinterpretation can be presented within a theoretical framework
based on the tripartite entanglement of {\it seven}
qubits \cite{Levfano,Ferrara}. The main idea is that $E_7$, as a
group of rank seven, contains seven copies of the rank-one groups
$SL(2)$. In quantum information, $SL(2)$ is the group of admissible
local manipulations of a qubit \cite{Dur}; this is the group of
stochastic local operations and classical communication (SLOCC).
Next, the fundamental $7\times 8=56$-dimensional irrep of $E_7$ can
be decomposed into seven copies of the $8$-dimensional three-qubit
Hilbert spaces according to a nice pattern dictated by the incidence structure of the Fano
plane \cite{BDL,LSV,Levfano,Geemen,Ferrara}. A similar seven-qubit
based understanding of the Lie-algebra of $E_7$ via the $133$-dimensional adjoint representation is also
possible \cite{Levfano,Manivel,Elduque}. Hence, within the context of the BHQC, a clear understanding of possible patterns of $SL(2)$-subgroups isomorphic to
$SL(2)^7$ of the $E_7$ is  of utmost importance. This task has been carried out by Cerchiai and van Geemen \cite{Geemen}.
Here, we would like to reiterate the basic idea of this work by showing its connection to the structure of our context space $\mathcal{I}$.

The basic observation of \cite{Geemen} is that the root lattice of $E_7$, $L(E_7)$, defined as
\beq
L(E_7)\equiv \{l_1\alpha_1+\dots +l_7\alpha_7\vert l_a\in\field{Z}\}
\eeq
\noindent
with $\alpha_a$, $a=1,2,\cdots, 7$, being the simple roots of $E_7$,
can be mapped into our vector space $V_3\simeq\field{Z}_2^6$ as follows
\beq
\pi:L(E_7)\to V_3,\qquad \pi(l_1\alpha_1+\dots +l_7\alpha_7)=l_1v_1+\dots +l_7v_7,
\eeq
\noindent
where the numbers $l_a$ on the right-hand side are to be understood mod $2$, and where the details of the correspondence between the simple roots $\alpha_a$, labeling the nodes of the Dynkin diagram of $E_7$, and certain three-qubit observables, $v_a$, can be found in \cite{Geemen}.
Here, we only note that the core of this correspondence is the relation
\beq
(\alpha_a,\alpha_b)=\langle v_a,v_b\rangle\quad {\rm mod}~2,
\label{mod2ident}
\eeq
\noindent
which establishes a relation between the inner product of the root system on the left-hand side and our symplectic product given by  Eq.\,(\ref{ezittaszimpl}) on the right-hand side.
Making use of $\pi$, one can map the $126$ roots of $E_7$ to the $63$ nonzero elements of $V_3$. Note that $\pi(\alpha)=\pi(-\alpha)$, and that the Weyl reflections in the root system correspond to the transvections of Eq.\,(\ref{transvections}).
Since the Weyl reflections generate $W(E_7)$ and the transvections generate $Sp(6,2)$, the map $\pi$ establishes the already-mentioned isomorphism $W(E_7)/{\field{Z}_2}\simeq Sp(6,2)$.

A positive root $\alpha$ induces an $sl(2)$-subalgebra with standard generators $\{X_{\alpha},X_{-\alpha},H_{\alpha}\}$, where $H_{\alpha}=[X_{\alpha},X_{-\alpha}]$ lies within the seven-dimensional Cartan subalgebra of $e_7$.
One can then show \cite{Geemen} that the generators of the subalgebras $sl_{\alpha}(2)$
and $sl_{\beta}(2)$, determined by two different positive roots $\alpha$ and $\beta$, commute if, and only if, these roots are orthogonal.
By virtue of Eq.\,(\ref{mod2ident}), this means that two commuting three-qubit observables can be associated with two commuting copies of $sl(2)$-algebras in $e_7$, i.\,e. with the SLOCC-algebras of two distinguishable qubits.
Since $E_7$ is of rank seven, its root system spans $\field{R}^7$; hence, there are no more than seven mutually commuting orthogonal roots. Using the map $\pi$,
this corresponds to the fact that the maximum number of pairwise commuting three-qubit observables is seven (our heptad).
Hence, the set of maximum sets of mutually orthogonal roots in the root system of $E_7$
has {\it the same} structure as our context space $\mathcal{I}$.
Moreover, since mutually orthogonal systems of roots correspond to an assignment of seven qubits with their seven commuting $sl(2)$ SLOCC-algebras, this establishes a correspondence between
our method(s) of studying $\mathcal{I}$ and the seven-qubit picture of the BHQC.
Indeed, as there are $135$ maximum sets of mutually orthogonal roots, there are also $135$ root subsystems $SL(2)^{\oplus 7}\subset E_7$ that can give rise to sets of seven-qubit systems occurring in the BHQC framework.

As a second application lying within the realm of the physics of
black holes in string theory, we will show that mapping sets of {\it
three-qubit} operators to a special subset of {\it four-qubit}
operators might even be relevant for shedding some new light on
issues of dimensional reduction within a finite geometric
framework. Since the main incentive of this paper is merely to set
the finite geometric ground for further applications in connection
with the BHQC, let us just present here several rudimentary observations
forming the basis of this interesting correspondence. The
detailed elaboration of these ideas will be postponed to a future
publication.

We have already mentioned at the beginning of this section that
the scalar fields play a crucial role in determining the
structure of {\it static} extremal black hole solutions of
effective four-dimensional supergravities. Mathematically, these fields
live in a symmetric space of the form $G_4/H_4$, where $G_4$ is
the four-dimensional $U$-duality group and $H_4$ its maximal
compact subgroup. In the case of maximal $N=8$ supergravity, we
have $G_4=E_{7(7)}$ and $H_4=SU(8)$. Working in the $SU(8)$-basis,
the central charge that shows up in the $E_{7(7)}$-symmetric
black-hole entropy formula is a complex antisymmetric $8\times 8$
matrix; this matrix can be expanded in terms of the $28$ antisymmetric
{\it three-qubit} operators as basis vectors, where the $56$ charges
are displayed as the complex expansion coefficients. In this
context, the relevant finite geometric structures have already been
discussed in detail \cite{LSV,LSVP}.
Moreover, the structure of the $E_{7(7)}$-symmetric black-hole
entropy formula can alternatively be described by the finite
geometry based on a particular graph studied by Cooperstein, which is closely related to the
$E_7$-Gosset polytope \cite{Coop}.

However, we can go even one step further and try to understand
{\it stationary} black-hole solutions. It is well known that
such solutions can effectively be studied by performing a
time-like {\it dimensional reduction} of the corresponding {\it
four}-dimensional supergravities, the result being a {\it
three}-dimensional gravity coupled to new scalars featuring a
nonlinear sigma model \cite{Breitenlohner}. These new scalar fields
that contain as a subset the original ones now form a
pseudo-Riemannian symmetric space $G_3/H_3^{\ast}$, with the line-element given by a pseudo-Riemannian metric. Here, $G_3$ is the
three-dimensional $U$-duality group and $H_3^{\ast}$ is the
maximally non-compact real form of $H_3(\mathbb{C})$, the
complexification of the maximal compact subgroup $H_3\subset G_3$.
In our special case, $G_3=E_{8(8)}$ and $H_3^{\ast}=SO^{\ast}(16)$.
It can be shown that in this picture extremal black-hole solutions
of the original four-dimensional theory can be mapped to the null
geodesics on the coset $G_3/H_3^{\ast}$ and are classified in
terms of the adjoint orbits under $G_3$ of a {\it nilpotent} $G_3$-Lie-algebra-valued conserved Noether charge
$Q$ \cite{Bossard1,Bossard2}. In addition to the usual electric and
magnetic charges (amenable to a finite geometric interpretation
based on three-qubit operators), $Q$ also contains the NUT-charge
and other conserved charges corresponding to four-dimensional duality rotations \cite{fake}. Hence, an attempt to understand these new quantities in a finite
geometric setting based on some $N$-qubit system, with $N>3$, is also very appealing.

Given our example of maximal supergravity, one can readily
characterize the $E_{8(8)}$-Lie-algebra-valued Noether charge in
an $SO^{\ast}(16)$-basis defined in terms of $16\times 16$
matrices . As the number of generators of this group is $120$, 
the relevant object serving as a suitable basis in this setting is a specific set of $120$ {\it four-qubit} operators. It associated finite geometric
structure can be found as follows.
The group $SO^{\ast}(16)$ is the matrix group of $16\times 16$
matrices $g$ satisfying \beq g^tg={\bf 1},\qquad g^{\dagger}\omega
g=\omega,\qquad \omega=Y\otimes I\otimes I\otimes I.
\label{sostar} \eeq \noindent Writing $g={\bf
1}+\mathcal{Z}+\ldots$, the above conditions  yield \beq
\mathcal{Z}^t=-\mathcal{Z},\qquad \mathcal{
Z}^{\dagger}\omega=-\omega\mathcal{Z}.\label{maskepp} \eeq
\noindent The latter equation implies that
$\omega\mathcal{Z}$ is a Hermitian matrix, whereas the former one leads to some extra constraints \beq
\omega\mathcal{Z}=S+iA,\qquad S^t=S,\quad S\omega=\omega S,\qquad
A^t=-A,\quad A\omega=-\omega A.\label{hermcond} \eeq \noindent Let
us now consider the quadratic form $Q_w$ (cf. Eq.\,(\ref{ujkvadforms})) associated with the element $w=(10001000)\in
V_4$ representing the four-qubit operator $\omega$. Using the
definition of $Q_w$ one can see that the points $v\in V_4$ of the
{\it elliptic} quadric $Q^-(7,2)$ given by the equation $Q_w(v)=0$ correspond to a set of
$119$ four-qubit operators.  According to Eq.\,(\ref{ujkvadforms}),
this set splits into $63$ symmetric operators that commute and $56$
antisymmetric ones that anticommute with $\omega$. Omitting the identity, 
these operators are precisely the ones occurring in the  expansion
of $\omega\mathcal{Z}$ (\ref{hermcond}), with $\mathcal{Z}$ being an element of the
Lie-algebra of $SO^{\ast}(16)$. Again disregarding the identity operator,
this expansion features altogether $63$ symmetric four-qubit
operators that lie on our quadric $Q^+(7,2)$ (see Sect.\,2). It is easy to see that
these $63$ operators are precisely the ones defined by
Eq.\,(\ref{hexapoints}), giving rise to the points of a copy of our
split Cayley hexagon. Hence, the finite geometric structures
studied in this paper occur naturally in this context of
time-like dimensional reduction.

On the other hand, the $120$ four-qubit operators contained in the
expansion
of $\omega\mathcal{Z}$ operators can be used to label the vertices of the
{\it projective version} of the $E_8$ Gosset
polytope \cite{Richter}. This is a finite geometric object on which the Weyl
group of $E_8$ -- a discrete subgroup of the three dimensional
$U$-duality group $G_3$ responsible for the 3D-version of
electric-magnetic duality -- acts naturally. Moreover, the $E_8$-polytope contains as the vertex figure the $E_7$ Gosset
polytope. In a natural labelling of the vertices of the $E_8$-polytope in terms of four-qubit operators, this $E_7$-polytope is
labelled by merely three-qubit ones. This subset, as expected,
gives back the graph of Cooperstein  encapsulating the structure
of the $E_{7(7)}$-symmetric black-hole entropy formula.

To briefly recapitulate, within the framework of the maximal $N=8$ supergravity, a  dimensional
 reduction from four to three dimensions can be characterized by particular mappings of three-qubit operators and
 their associated finite geometric structures to certain sets of
 four-qubit operators with their corresponding finite geometric
 objects. These mappings and allied structures can possibly provide a finite
 geometric way for understanding the structure of the Noether
 charge $Q$ and, via the nilpotent orbits of $Q$, also of the different
 classes of the extremal black-hole solutions.

We mention in closing that the physical meaning of the extra qubit
showing up under dimensional reduction can nicely be identified in
the so-called STU truncation \cite{Levay4qbit} of $N=8$
supergravity, where the extra $SL(2)$ group acting on the extra
qubit corresponds to the Ehlers group \cite{Ehlers} well known to
general relativists. A similar identification of physical
quantities on the black-hole side with the ones on the finite-geometric side is envisaged. 
In a future work, we will hopefully be
able to provide the reader with a more mature and elaborated form of this
correspondence.

\section{Conclusions}
We have gained substantial insights into a yet-unnoticed relation between the three-qubit and four-qubit generalized Pauli groups, based on the so-called spin-module of the symplectic group $Sp(6,2$). Our starting point was the set $\mathcal{I}$ of 135 heptads of pairwise commuting three-qubit observables. We first labeled the elements of this distinguished  subspace of $\mathcal{W}(5,2)$ by those of a seven-dimensional Clifford algebra. Then, by employing the formalism of Pl\"ucker/Grassmann embeddings, we worked out an explicit form of the bijection between  $\mathcal{I}$ and the set of 135 symmetric four-qubit observables, lying on a particular hyperbolic quadric of $\mathcal{W}(7,2)$. After performing a detailed analysis of the action of $Sp(6,2)$ on both sides of
this correspondence, we gave a couple of interesting physical applications of our formalism. The first application concerned the structure of the set of 12096 Mermin's pentagrams living
in $\mathcal{W}(5,2)$, as recently discovered with the aid of a computer \cite{Holweck}. Here, we have not only succeeded in furnishing a rigorous, computer-free explanation why there exist just 336 such pentagrams formed from the symmetric three-qubit observables, but also shown that the whole set of these ``symmetric'' pentagrams can be generated from merely {\it six} basic types (see Eqs.\,(\ref{antiflag1})--(\ref{quad2})). Moreover, we also offered some hints --- linked with the structure of the split Cayley hexagon of order two --- towards accounting for the number 12096 as well. Our second, BHQC, application made use of the fact that $Sp(6,2)$
is related to the Weyl-group of $E_7$, which is a subgroup of the
$U$-duality group responsible for electric-magnetic duality, which thus enabled
us to reveal the relevance of our formalism for the corresponding BHQC framework.

\section{Acknowledgments}
A major part of this research was conducted within the ``Research in Pairs'' program of the Mathematisches Forschungsinstitut Oberwolfach (Oberwolfach, Germany), in the period from 24 February 
to 16 March, 2013. PL would also like to acknowledge financial support he received from the MTA-BME Condensed Matter Physics Research Group, grant No.\,04119. MS was also partially supported by the VEGA Grant Agency, grant No.\,2/0003/13. We also thank the anonymous referee for a number of constructive remarks.


\begin{thebibliography}{}
\bibitem{Vrana} The explicit form of this representation for the generator $\beta$ and a presentation for $G_2(2)$ was determined by P. Vrana in his MSc Thesis.
\end{thebibliography}
\end{document}